\documentclass{aa}
\usepackage{graphicx,txfonts}
\usepackage{natbib}

\newdimen\captwidth
\newdimen\figwidth

\newcommand{\mjup}{M_\mathrm{Jup}}

\newcommand{\excs}{\extracolsep{\fill}}

\begin{document}

\title{Insights on the dynamical history of the Fomalhaut system} 
\subtitle{Investigating the Fom c hypothesis}
\author{V. Faramaz$^{1,2}$ \and H. Beust$^{1,2}$ \and J.-C. Augereau$^{1,2}$ \and P. Kalas$^3$ \and J.R. Graham$^3$}
\institute{Univ. Grenoble Alpes, IPAG, F-38000 Grenoble, France \and CNRS, IPAG, F-38000 Grenoble, France \and Department of Astronomy, 601 Campbell Hall, University of California at Berkeley, Berkeley, CA 94720, USA}
\date{Accepted for publication in A\&A}
\offprints{V. Faramaz}
\mail{Virginie.Faramaz@obs.ujf-grenoble.fr}
\titlerunning{The Fomalhaut c hypothesis}
\authorrunning{Faramaz et al.}
\abstract
{The eccentric shape of the debris disk observed around the star Fomalhaut was first attributed to Fom b, a companion detected near the belt inner-edge, but new constraints on its orbit revealed that it is belt-crossing, highly eccentric $(e \sim 0.6-0.9)$, and can hardly account for the shape of the belt. The best scenario to explain this paradox is that there is another massive body in this system, Fom c, which drives the debris disk shape. The resulting planetary system is highly unstable, which hints at a dynamical scenario involving a recent scattering of Fom b on its current orbit, potentially with the putative Fom c.}
{Our goal is to give insights on the probability for Fom b to have been set on its highly eccentric orbit by a close-encounter with the putative Fom c. We aim to study in particular the part played by mean-motion resonances with Fom c, which could have brought Fom b sufficiently close to Fom c for it to be scattered on its current orbit, but also delay this scattering event.}
{We assumed that Fom c is much more massive than Fom b, that is, Fom b behaves as a mass-less test-particle compared to Fom c. This allowed us to use N-body numerical simulations and to study the influence of a fixed orbit Fom c on a population of mass-less test-particles, that is, to study the generation of Fom b-like orbits by direct scattering events or via mean-motion resonance processes. We assumed that Fom b originated from an orbit inner to that of the putative Fom c.}
{We found that the generation of orbits similar to that of Fom b, either in term of dimensions or orientation, is a robust process involving a scattering event and a further secular evolution of inner material with an eccentric massive body such as the putative Fom c. We found in particular that mean-motion resonances can delay scattering events, and thus the production of Fom b-like orbits, on timescales comparable to the age of the system, thus explaining the witnessing of an unstable configuration.}
{We conclude that Fom b probably originated from an inner resonance with Fom c, which is at least Neptune-Saturn size, and was set on its current orbit by a scattering event with Fom c. Since Fom b could not have formed from material in resonance, our scenario also hints at former migration processes in this planetary system.}
  \keywords{Stars: Fomalhaut -- Planetary systems -- Circumstellar matter 
  -- Methods: numerical -- Celestial mechanics}
\maketitle

\section{Introduction}
Fomalhaut A ($\alpha\:$Psa) is a 440 Myr old \citep{2012ApJ...754L..20M} A3V star, located at 7.7\,pc \citep{2007ASSL..350.....V,2012ApJ...754L..20M}.  
As revealed by HST, Fomalhaut A is surrounded by an eccentric dust ring ($\mathrm{e}=0.11\pm 0.01$) with a sharp inner edge at 133\,AU and extending up to 158\,AU \citep{2005Natur.435.1067K}. This eccentric shape hinted at the presence of a massive body orbiting inside the belt on an eccentric orbit, dynamically shaping the belt \citep{2006MNRAS.372L..14Q,2005ApJ...625..398D}. This hypothesis was furthermore confirmed by the direct detection of a companion near the inner edge of the belt, Fomalhaut~b (hereafter Fom b) \citep{2008Sci...322.1345K}. 
The nature of Fom b has been intensely discussed since its discovery because it is observed at visible wavelengths, but remain undetected in the infrared \citep{2008Sci...322.1345K,2009ApJ...700.1647M,2012ApJ...747..116J}. The consensus today is that it is a planetary body, surrounded by a population of dust, either in the form of a planetary ring system \citep{2008Sci...322.1345K}, or a dust cloud resulting from a collision between satellites \citep{2011MNRAS.412.2137K,2014ApJ...786...70K}. Further observations of this body led to the detection of its orbital motion. 
Based on the available astrometric points, the first attempts to constrain its orbit showed that it is surprisingly extremely eccentric, nearly coplanar and close to apsidal alignment with the belt, so that the orbit inevitably crosses it \citep[$\mathrm{a_b}\sim 110-120\,$AU and $\mathrm{e_b} \simeq 0.92-0.94$,][]{2013AAS...22132403G,2014A&A...561A..43B}. 
Detailed recent dynamical investigations \citep{2014A&A...561A..43B} revealed that this orbital configuration is not compatible with the shape of the disk and the age of the system. A low eccentricity belt like the one observed might indeed be produced by this perturber, but irrespective of the mass of Fom b, it appears to be a transient feature that evolves to very high eccentricities and possibly to its destruction on timescales much smaller than the age of the system.

Since Fomalhaut A is part of a triple star system, a possible explanation could be that the disk is shaped by another stellar component. A dynamical study by \citet{2014MNRAS.442..142S} on interactions between the debris disk of Fomalhaut A and other members of this triple star system shows that an approach of Fomalhaut C could excite the disk eccentricity to the observed value, without need of a planet to account for it. However, Fomalhaut C is part of the widest known stellar companions, with a semi-major axis estimated to $a_C \sim 0.5\,$pc$\sim 200\,$kAU \citep{2013AJ....146..154M}. Since the secular timescale scales as $\alpha^{-3}$, where $\alpha=a_d/a_C$ is the ratio of the semi-major axes of a disk and its perturber, here Fomalhaut C \citep[see for instance Eq. 6 of ][]{2014A&A...563A..72F}, with a disk of semi-major axis $a_d \sim 100\,$AU, the secular timescale is of Gyr order. 
As a consequence, the secular action of Fomalhaut C that produces the eccentric disk would require timescales much greater than the age of the system to take place.

In the end, the most straightforward solution to this apparent paradox is to suppose the presence of a second more massive and yet undetected body in the system (hereafter named Fom c), which is responsible for the disk shaping because of a predominant dynamical influence.  This implies that Fom b is rather a low-mass body compared to the putative Fom c, but other arguments suggest this. As shown by \citet{2014A&A...561A..43B}, even with no Fom c and given its orbit, a massive Fom b would lead to a rapid destruction of the observed belt. Moreover, if Fom b was massive enough, it would trigger a more or less rapid secular orbital precession of the orbit of Fom c. This could prevent Fom c from sustaining the belt asymmetry. A dynamical study of \citet{2014MNRAS.438.3577T} also suggests that the best scenario that matches the observational constraints is that of a super-Earth Fom b with an undetected belt-shaping Saturn-mass planet.
Finally, a low-mass Fom b is also compatible with recent photometric studies, which suggest that it is no more than Earth- or Super-Earth sized \citep{2012ApJ...747..116J,2013ApJ...769...42G}.

Considering that the orbit of Fom b is highly eccentric with an apastron beyond the outer edge of the belt and a periastron that could be as low as a few AU \citep{2014A&A...561A..43B}, and that the putative Fom c would move on a less eccentric orbit located slightly inside the inner edge of the belt, then inevitably both orbits are expected to cross each other. This raises the question of the dynamical stability of Fom b. In this scenario, its present day orbit could just be a transient configuration. It could have been put there by a more or less recent scattering event, potentially with Fom c \citep{2014A&A...561A..43B}, and could be subject to an ejection in a more or less near future. The more massive is Fom c, the shorter the survival timescale of Fom b. In \citet{2014A&A...561A..43B}, it is argued that Fom c should probably be $\sim$Saturn-sized to be able to shape the belt while not ejecting Fom b too quickly from its present-day orbit, just to make it likely for us to witness the transient configuration today.

The goal of this paper is to investigate the issues of the generation of the present-day orbital configuration of Fom b. In this work, we discuss whether models involving Fom c only can explain both the orbit of Fom b and the shape of the outer Kuiper-belt. We examinate how Fom b, starting from a configuration inner to Fom c could have been put on its present-day orbit by a scattering event with Fom c. We show that mean-motion resonances may play a crucial role by delaying this scattering event. We outline our method and our expectations in Sect.~\ref{sec:method}, and display our results in Sect.~\ref{sec:results}. We discuss these results in Sect.~\ref{sec:discussion}. In particular, we investigate the influence of the eccentricity and the mass of Fom c, which reveal to be crucial parameters controlling the mechanism that generates orbits comparable to that of Fom b. The mechanism itself is also shown to be more complex than originally thought and is investigated in more details. We give our conclusions in Sect.~\ref{sec:conclusion}.


\section{Method}\label{sec:method}
\begin{figure}
\resizebox{\hsize}{!}{\includegraphics[width=0.55\textwidth,height=0.45\textwidth]{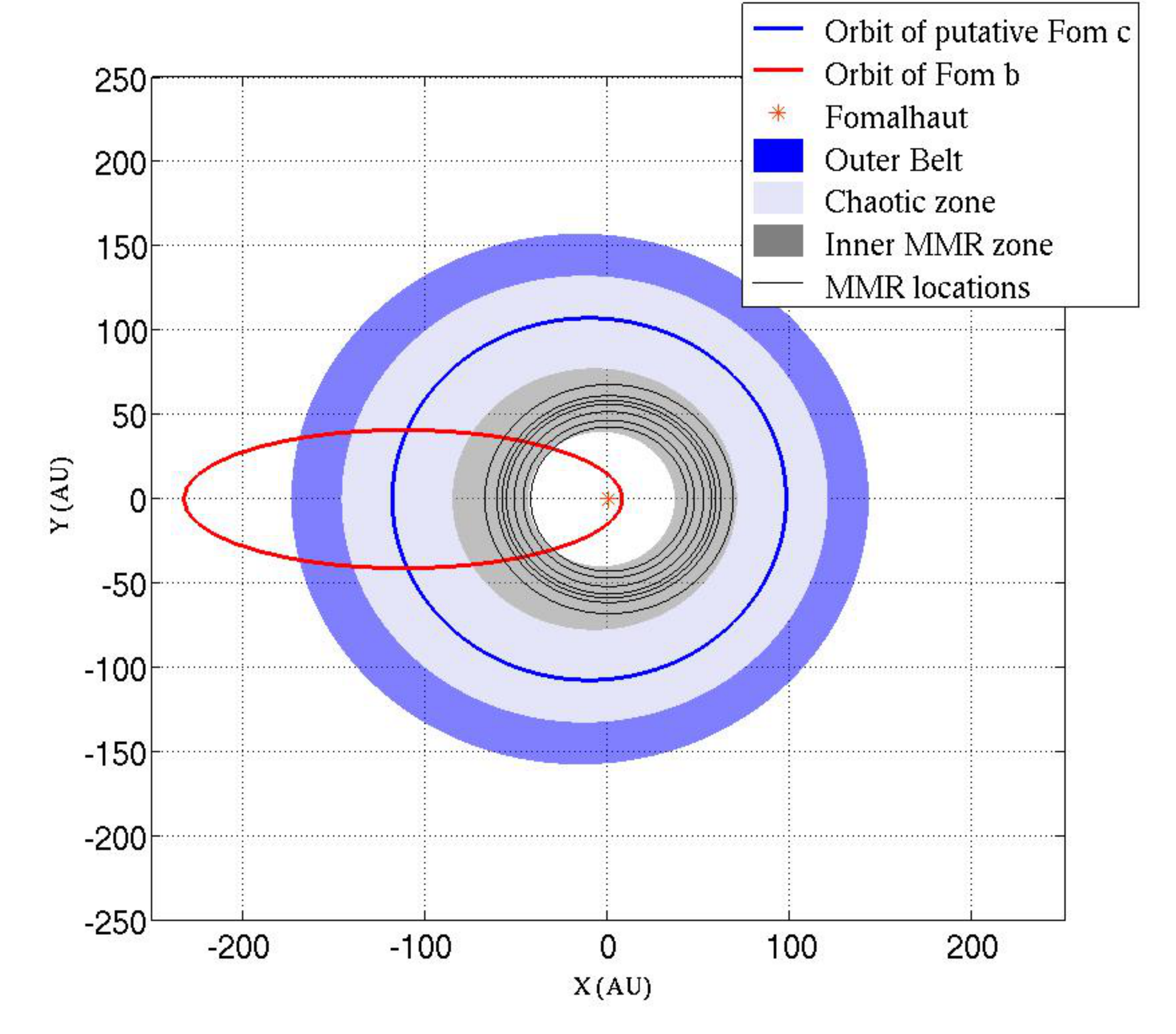}}
\caption[]{Schematic configuration of the Fomalhaut system. The outer Kuiper belt extends from 133 to 158 AU and has an eccentricity of 0.1. The orbit of Fom b has $a_b=120\,$AU and $e_b=0.94$, that is, the peak values derived by \citet{2014A&A...561A..43B}. Since Fom b has a low inclination relative to the outer belt and is nearly apsidally aligned with it, we represent this orbit as coplanar and apsidally aligned with the belt for sake of simplicity. We represent as well the presumed orbit of the putative Fom c, and the regions of potential origin for Fom b investigated in this paper, that is, the chaotic zone of Fom c and the region where mean-motion resonances which may generate orbits crossing the chaotic zone of Fom c are localised.}
\label{fig:whole_system}
\end{figure}

The basic assumption of our study is that Fom c is significantly more massive that Fom b. This is supported by a recent study \citep{2014A&A...561A..43B} showing that Fom b is probably a low-mass object and that the eccentric disk shape is controlled by another, more massive object, presumably Fom c. We use N-body simulations to investigate the ability of the putative Fom c to put Fom b on its present-day orbit and the typical timescale for this to happen starting from various initial configurations. Thanks to the mass difference between both objects, Fom b will be treated in this work as a massless test-particle perturbed by Fom c. Our second assumption is that Fom b was originally on an orbit inner to that of Fom c. The configuration of the system is illustrated in Fig.~\ref{fig:whole_system}. 
  
In this section, we present our numerical set-up as well as theoretical background on the production of Fom b-like orbits, either by a direct scattering event or via mean-motion resonances.

\subsection{The putative Fom c}
The numerical set-up of the orbital parameters and the mass of the putative Fom c was chosen considering constraints on the belt-shepherding planet retrieved from previous dynamical studies and observational constraints.

Based on dynamical investigations, \citet{2009ApJ...693..734C} found an upper-mass limit of $3\,\mathrm{M_{Jup}}$, a lower semi-major axis limit of $\sim 101.5\,$AU, and an orbital eccentricity of $\sim 0.11-0.13$, while \citet{2006MNRAS.372L..14Q} found that the belt-shaping planet should rather be Neptune or Saturn-mass, with a semi-major axis of $\sim 120\,$AU and an orbital eccentricy of $\sim 0.1$. On the other hand, \citet{2013arXiv1311.1207R} found a mass of $\sim 7.6\pm 4.6\,\mathrm{M_{Jup}}$, a lower semi-major axis limit of $\sim 85\,$AU, and an orbital eccentricity of $\sim 0.11$. Observational constraints from \citet{2012ApJ...747..116J} and \citet{2013ApJ...777L...6C} gave a detection limit of $\sim 1-3\,\mathrm{M_{Jup}}$ at these distances.

Therefore, we arbitrarily chose a mass of $\mathrm{m_c}=3 \,\mathrm{M_{Jup}}$ for Fom c and will discuss the impact of this mass on our results in Sect.~\ref{sec:discussion}. We followed the approach of \citet{2006MNRAS.372L..14Q,2009ApJ...693..734C} and \citet{1989Icar...82..402D}, and considered that the belt inner edge, located at 133 AU, was created by the chaotic zone around Fom c, that is, the region where material is on a highly unstable orbit. In this context, the location of the inner edge should correspond to the outer boundary of the chaotic zone of Fom c. The chaotic zone of a planet is defined as the region where mean-motion resonances overlap. The width of this zone, $\mathrm{\Delta a}$, depends on the mass of the planet, $\mathrm{m_{planet}}$, and its semi-major axis $\mathrm{a_{planet}}$. It reads : 
\begin{equation}
\mathrm{\frac{\Delta a}{a_{\mathrm{planet}}} = \frac{a_{\mathrm{edge}}-a_{\mathrm{planet}}}{a_{\mathrm{planet}}} = 1.5 \mu^{2/7}}\qquad,
\end{equation}
where $\mu = \mathrm{m_{planet}/M_\star}$ \citep{1980AJ.....85.1122W,1989Icar...82..402D}. The mass of Fomalhaut was set to $\mathrm{M}_{\star}=1.92 \,\mathrm{M}_{\odot}$ \citep{2012ApJ...754L..20M}.  Consequently, one can deduce the semi-major axis of a planet of a given mass that generates a disk inner edge at 133 AU: 

\begin{equation}\label{eq:semi_maj}
\mathrm{a_{\mathrm{planet}} = \frac{a_{\mathrm{edge}}}{1+1.5\mu^{2/7}}}\qquad.
\end{equation}

We derived a semi-major axis of $\mathrm{a_c}=107.8\,$AU. We set the orbital eccentricity of Fom c to be $\mathrm{e_c}=0.1$ as to match the observed ellipticity of the belt. Indeed, setting the eccentricity of the planet to that of the belt
is appropriate in the limit of the planet being close to the belt, coplanar with the belt(or close to coplanarity), and a single planet being responsible for the forced eccentricity of the belt, as shown by the pericenter-glow model of \citet{1999ApJ...527..918W}.

This is in agreement with other studies which showed that the chaotic zone extends radially up to $\sim 3-3.5\mathrm{R_H}$ around the planet's orbit \citep{2000ApJ...534..428I,2009Icar..199..197K}, where $\mathrm{R_H}$ is the Hill radius defined by

\begin{equation}\label{eq:rhill}
\mathrm{R_H= a_c \left(\frac{m_c}{3\mathrm{M}_{\star}} \right)^{1/3}} \qquad.
\end{equation}

Eq.~\ref{eq:rhill} gives then $\mathrm{R_H}=8.54\,$AU, and with $\mathrm{a_c}=107.8\,$AU, the inner edge of the belt at 133 AU is indeed distant of $\sim 3\mathrm{R_H}$ from Fom c.
However, this approach holds for circular planets, and if we chose to follow this one, in accordance to previous studies on the shaping of the outer belt made by \citet{2006MNRAS.372L..14Q,2009ApJ...693..734C}, we shall see in Sect.~\ref{sec:results} that the width of the gap actually slightly differs in the case of an eccentric planet, as unraveled by \citet{2012MNRAS.419.3074M}. This means that the semi-major axis of Fom c derived according to the approach of \citet{2012MNRAS.419.3074M} will slightly differ from the one we have derived, but nevertheless does not impact the mechanism which can set inner particles on orbits similar to this of Fom b.

\subsection{Producing Fom b-like orbits from originally quiescent orbits}

Driving Fom b from an orbit located inside this of Fom c to its present-day orbit means drastically increasing its semi-major axis and its eccentricity. The most straightforward mechanism to achieve this is a scattering event caused by a close encounter. Such events are expected to occur in the chaotic zone around Fom c outlined above. Only a close encounter event can induce sudden changes of the eccentricity and semi-major axis of a test-particle, potentially in a way that sets it on a Fom b-like orbit. 

To undergo a scattering event, a particle must cross the chaotic zone at some point on its orbit. It can have formed there or have formed more deeply inside and then have suffered an orbital evolution that caused its orbit to furthermore cross the chaotic zone. Particles moving initially in the chaotic zone have very few chances to survive there more than a few $10^7\,$yrs, i.e. much less than the age of the star. Hence we think that models involving a Fom b that formed deeper inside the orbit of Fom c and was furthermore driven to cross the chaotic zone are more suited to our purpose.

To lead an inner orbit to cross the chaotic zone, its semi-major axis and/or its eccentricity must be increased. A major semi-major axis change can only be achieved by a scattering event with other unknown planets prior to crossing the chaotic zone of Fom c. Of course, this cannot be excluded, but implies the hypothetical presence of at least a third planet, with potentially similar timescale problems as with close encounters with Fom c. Here we will investigate models involving Fom c only.

For a low-mass Fom b forming inside the orbit of Fom c on a low eccentricity orbit, the only way to make it reach the chaotic zone with no scattering event by another inner planet is to drastically increase its eccentricity by trapping Fom b in near mean-motion resonance with Fom c. Regular secular perturbations triggered by Fom c on particles moving deeply inside its orbit are indeed known not to much affect their semi-major axes.  Moreover, particles moving outside resonances are expected to undergo only moderate amplitude regular eccentricity fluctuations. Conversely, particles trapped in some mean-motion resonances can sometimes see their eccentricity increased to very large values and therefore constitute valuable routes to cause a Fom b progenitor to cross the chaotic zone of Fom c.

Mean motion resonances (hereafter MMR) between a particle and a perturber (here Fom c), usually noted $\mathrm{n:p}$, where $\mathrm{n}$ and $\mathrm{p}$ are integers, concern particles with orbital periods achieving the $\mathrm{p/n}$ commensurability with that of the perturber. The integer $\mathrm{q=|n-p|}$ is called the order of the resonance. MMRs occur at specific locations relative to the orbit of the perturber. Resonances with $\mathrm{n>p}$ correspond to \emph{inner} resonances, that is, particles orbiting inside the orbit of the perturber, while $\mathrm{n<p}$ denotes \emph{outer} resonances.  

Particles trapped in MMRs are characterized by the libration of a characteristic resonant angle $\sigma$ \citep[see][for details]{1996Icar..120..358B,1995Icar..115...60M} and small amplitude semi-major axis librations ($\la 0.1$AU) around the exact resonance location. If the eccentricity of the perturber is zero (or very small), then the secular evolution of the eccentricity is coupled with that of the semi-major axis, so that the eccentricity only undergoes small amplitude variations. But if the eccentricity of the perturber is non-zero, the eccentricity modulations can have much larger amplitudes. \citet{1989A&A...213..436Y} showed that this is particularly relevant for inner MMRs like the 4:1, 3:1 or 5:2. This mechanism is thought to be responsible for the generation of the Kirkwood gaps in the solar system \citep{1983Icar...56...51W}, and it has been claimed to trigger the Falling Evaporating Bodies (FEBs, that is, star-grazing evaporating planetesimals, or comets) mechanism towards $\beta\:$Pictoris \citep{1996Icar..120..358B,2000Icar..143..170B}. As soon as the eccentricity of the perturber overcomes $\sim0.05$, this mechanism is able to increase the eccentricity of resonant particles up to large values in the cases of the 3:1, 5:2, 7:3 and 2:1 resonances, and even virtually $\sim 1$ in the case of the 4:1 resonance. In the asteroid belt, other resonances can also be active provided it overlaps with the $\nu_6$ secular resonance \citep{1989A&A...213..436Y}. Moreover, as we shall see in Sect.~\ref{sec:results}, material in MMRs need some time to reach the eccentricity required to cross the chaotic zone, which can thus delay a scattering event. If this delay were to be comparable to the age of the system, this would explain why the unstable present-day configuration is witnessed.

\subsection{Initial sets of particles}
In order to investigate the scenarios outlined above, we ran several simulations with different initial sets of particles, that is, potential Fom b planets, each of them corresponding to either a specific MMR or a wide range of semi-major axes including both MMRs and the chaotic zone relative to Fom c. The initial conditions of our simulations are all summarised in Table~\ref{tab:initial}. Each of these test-particles populations are assumed to be coplanar with the orbit of Fom c. Orbital inclinations with respect to the orbit of Fom c orbit were randomly distributed between 0 and $3\degr$, while the eccentricities were distributed between 0 and 0.05. All remaining initial angles, that is, the longitude of periastron, the longitude of ascending node, and initial mean anomaly, were randomly drawn between 0 and $2\pi$ in a uniform way.

For each simulation, the initial semi-major axes of the test-particles were also uniformly and randomly distributed between boundaries that were specific to each of them.

\begin{table}[htbp]
\caption{Characteristics of all initial sets of particles used in our
  numerical study. In all runs, Fom c itself is assumed to be a
  $3\,\mathrm{M_{Jup}}$ planet orbiting Fomalhaut with semi-major axis
  $\mathrm{a_c}=107.8\,$AU and eccentricity $\mathrm{e_c}=0.1$. 
   All sets of particles are ring-like belts of test-particles extending
  radially between boundaries given below, eccentricities randomly
  chosen between 0 and 0.05, and inclinations between 0 and $3\degr$
  relative to Fom c's orbital plane. Run~A contains 
  250,000 particles and runs~B--H contain 100,000 particles.} 
\label{tab:initial}
\begin{tabular*}{\columnwidth}{@{\excs}llll}
\hline\hline\noalign{\smallskip}
Run \# & Dynamical status & Semi-major axis & Theoretical resonance\\
       & relative to Fom c & extent $\mathrm{a}$ (AU) & location $\mathrm{a_{n:p}}$ (AU) \\
\noalign{\smallskip}
\hline\noalign{\smallskip}
A & Broad distribution & 40-110 & - \\
B & 4:1 MMR & 40.3--45.3 & 42.8 \\
C & 7:2 MMR & 44.3--49.3 & 46.8 \\
D & 3:1 MMR & 49.3--54.3 & 51.8 \\
E & 8:3 MMR & 53.6--58.6 & 56.1 \\
F & 5:2 MMR & 56.0--61.0 & 58.5 \\
G & 7:3 MMR & 58.8--63.8 & 61.3 \\
H & 2:1 MMR & 65.4--70.4 & 67.9 \\
\noalign{\smallskip}\hline
\end{tabular*}
\end{table}

Run~A is dedicated to study the dynamics of 250,000 particles widely distributed radially, which covers both the chaotic zone and the locations of the MMRs, for comparison. The initial semi-major axes of the test-particles were distributed randomly between $[40\,\mathrm{AU};110\,\mathrm{AU}]$. Here the upper limit corresponds to the apastron of Fom c minus one Hill radii, to take into account the eccentricity of Fom c, and the lower limit extends the distribution of the test-particles slightly further in than the 4:1 MMR.
Runs~B--H from Table~\ref{tab:initial} focus on rings of 100,000 test-particles centered on specific MMRs with Fom c. Not all MMRs needed actually to be tested. As long a they keep trapped in a MMR, the  semi-major axes of test-particles do not vary significantly, as they only undergo small amplitude secular variations around the theoretical MMR value (see Table~\ref{tab:initial}). This remains true even as their eccentricity approaches 1. Therefore, their apastron cannot grow higher than twice the theoretical $\mathrm{a_{n:p}}$ value, and we limited ourselves to MMRs achieving this condition.
Note that the further in the MMR is located, the higher the eccentricity of a test-particles should increase in order for its orbit to cross the chaotic zone. Therefore, it is expected in a general manner that the most inner MMRs such as the 3:1 and 4:1 should be less efficient routes to generate orbits comparable to that of Fom b in our scenario. 


The ability of a MMR to set a test-particle on a orbit sufficiently eccentric to cross the chaotic zone, or even the orbit of Fom c, can be evaluated thanks to phase-space diagrams. 

The theory of resonant phase-space diagrams is outlined in 
\citet{1996Icar..120..358B}. In the framework of the restricted three-body 
system, the interaction Hamiltonian $H$ acting on Fom~b in 
stellocentric refrence frame reads
\begin{equation}
H=-\frac{GM}{2a}
-Gm_c\left(\frac{1}{\left|\vec{r}-\vec{r_c}\right|}
-\frac{\vec{r}\cdot\vec{r_c}}{r_c^3}\right)\qquad,
\label{hamil0}
\end{equation}
where $M$ is the total mass of the system (nearly equal to Fomalhaut's mass),
$G$ is the gravitational constant, $\vec{r}$ and $\vec{r_c}$ are the position vectors of Fom b and c in heliocentric frame respectively, and $a$ is the semi-major axis of Fom b. We shall restrict ourselves to the planar
problem, where all three bodies move in the same plane. Note that when inclinations are small (which is supposed here), this is still valid for describing the secular motion as far as semi-major axes and eccentricities are concerned. 
With this assumption, the Hamiltonian $H$ reduces to two degrees of freedom. As explained in \citet{2014A&A...561A..43B}, if the two planets are not locked in a MMR, the secular motion can be accurately described taking the time average of Hamiltonian $H$, independently over both orbits. This is of course not the case here. As explained in \citet{1996Icar..120..358B}, a canonical transformation must be performed first. Let us consider that Fom~b is locked in a $p+q:p$ MMR with respect to Fom~c. The transformation is made introducing the variables
\begin{eqnarray}
\sigma & = & \frac{p+q}{q}\lambda_c-\frac{p}{q}\lambda-\varpi\quad;\\
\nu & = & \varpi-\varpi_c\quad;\\
N & = & \sqrt{aGM}\left(\sqrt{1-e^2}-\frac{p+q}{p}\right)\quad,
\end{eqnarray}
where $\lambda$ and $\lambda_c$ are the mean longitudes of Fom~b and Fom~c, and $\varpi$ and $\varpi_c$ their longitudes of periastra, and $e$ is the eccentricity of Fom~b's. The variable $\sigma$ is usally called the ''cricital angle of resonance''. It is a slowly varying quantity thanks to the MMR. The variable $\nu$ is the longitude of periastron of Fom b with respect to that of Fom~c. The canonical transformation induces a change of Hamiltonian, because it implicitly depends on time through $\lambda_c$, so that the new Hamiltonian reads
\begin{equation}
H'=H-\frac{p+q}{q}n_c\sqrt{a_bGM}\qquad,
\end{equation}
where $n_c$ is the mean motion of  Fom~c.

Non-resonant orbits are characterised by a more or less regular circulation of $\sigma$, while resonant ones exhibit a libration of $\sigma$ around an equilibrium value. The $\sigma$-libration motion induces eccentricity oscillations. 
Whenever the perturbing planet (Fom~c here) moves on a circular orbit, then the Hamiltonian $H'$ does no longer depend on $\nu$, and variable $N$ turns out to be a secular constant of motion. Hence eccentricity oscillations also trigger coupled semi-major oscillations to keep $N$ constant.

When the orbit of the perturbing planet is not circular but still moderately eccentric, then $N$ is no longer preserved, but the $(a,e)$ libration motion still applies. The value of $N$ is subject to a slow drift. This can drive the eccentricity to very high values in some cases \citep{1996Icar..120..358B,1995Icar..115...60M}. This is exactly the purpose of our model here.

\begin{figure*}[!t]
\makebox[\textwidth]{\includegraphics[width=0.3\textwidth,height=0.3\textwidth]{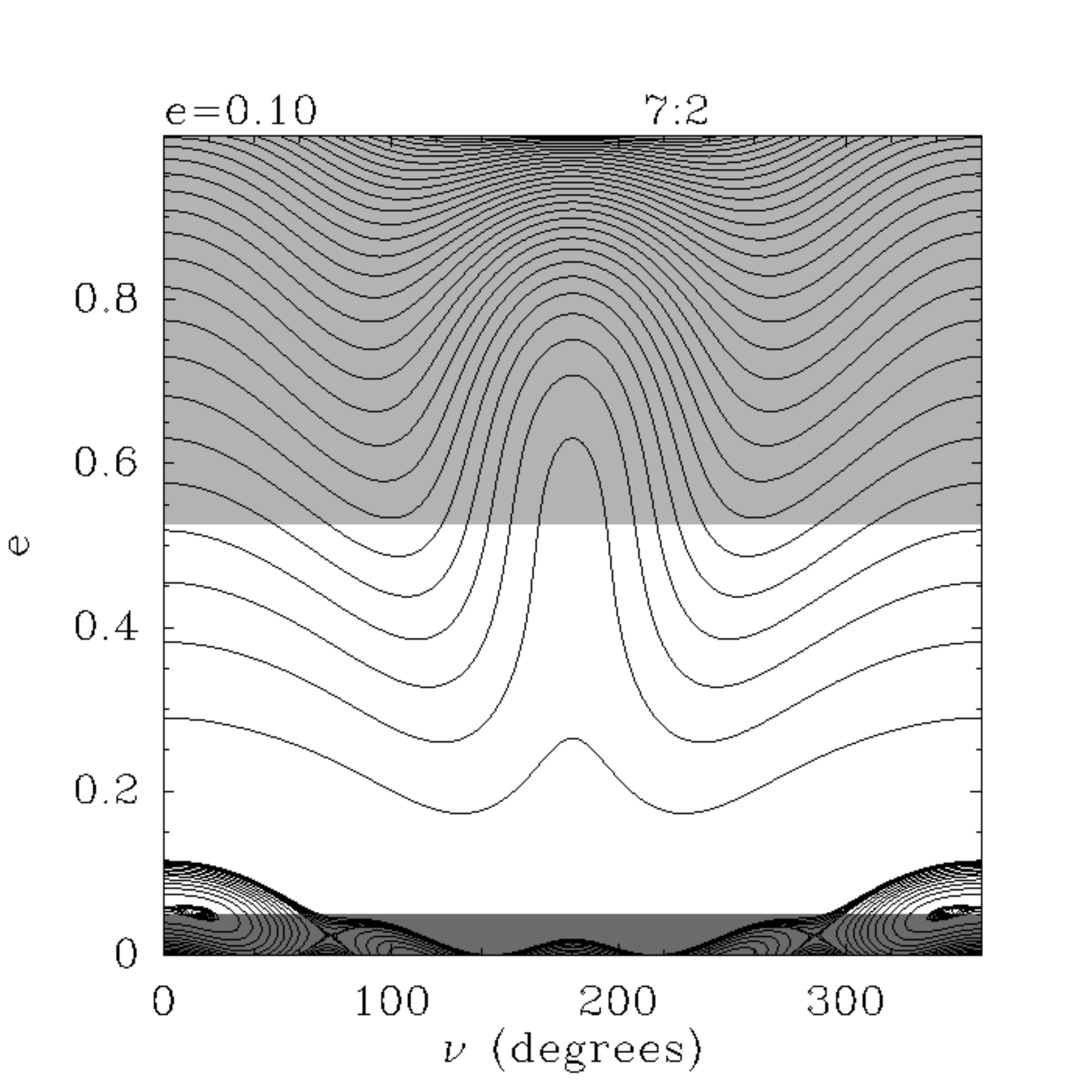}
\includegraphics[width=0.3\textwidth,height=0.3\textwidth]{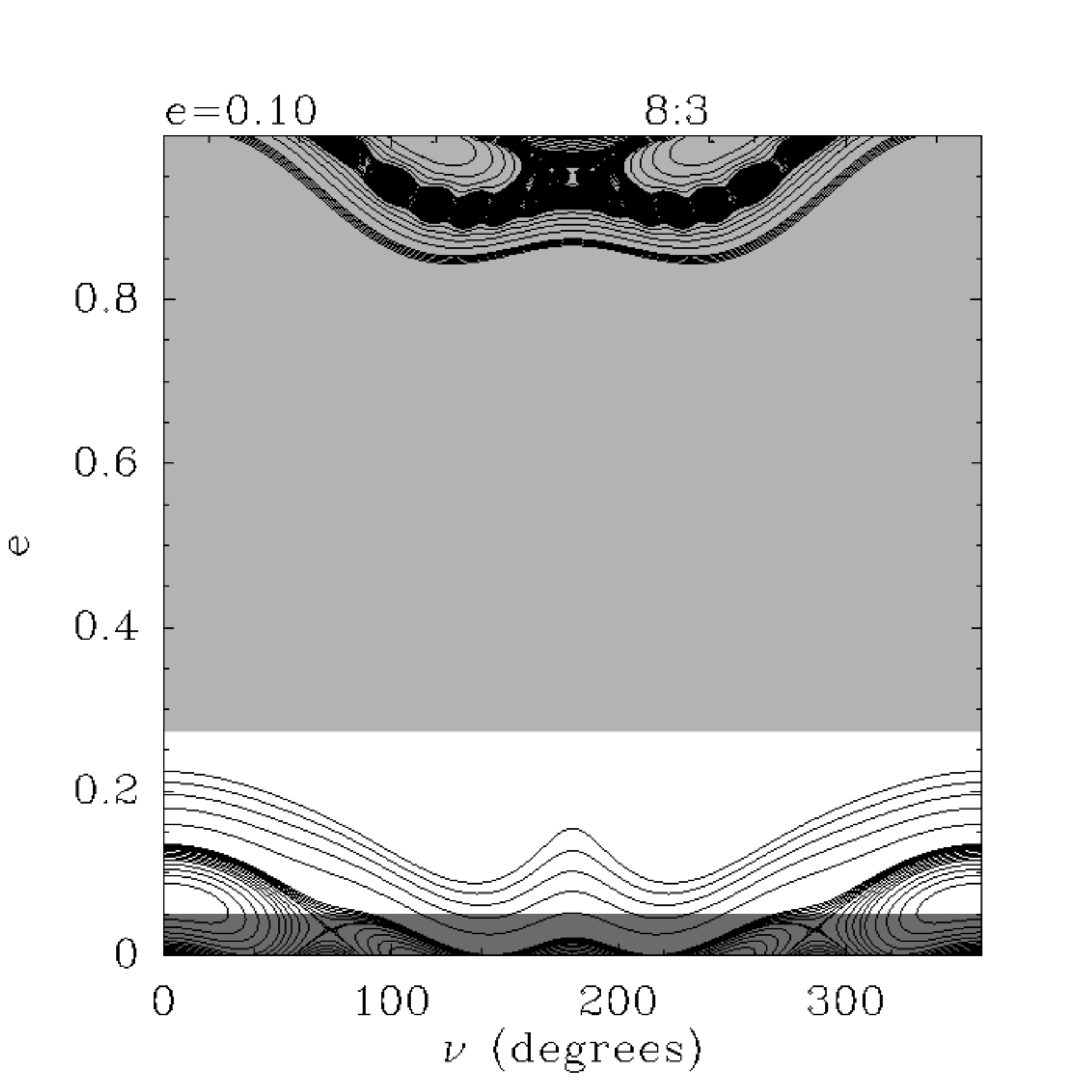}}
\makebox[\textwidth]{\includegraphics[width=0.3\textwidth,height=0.3\textwidth]{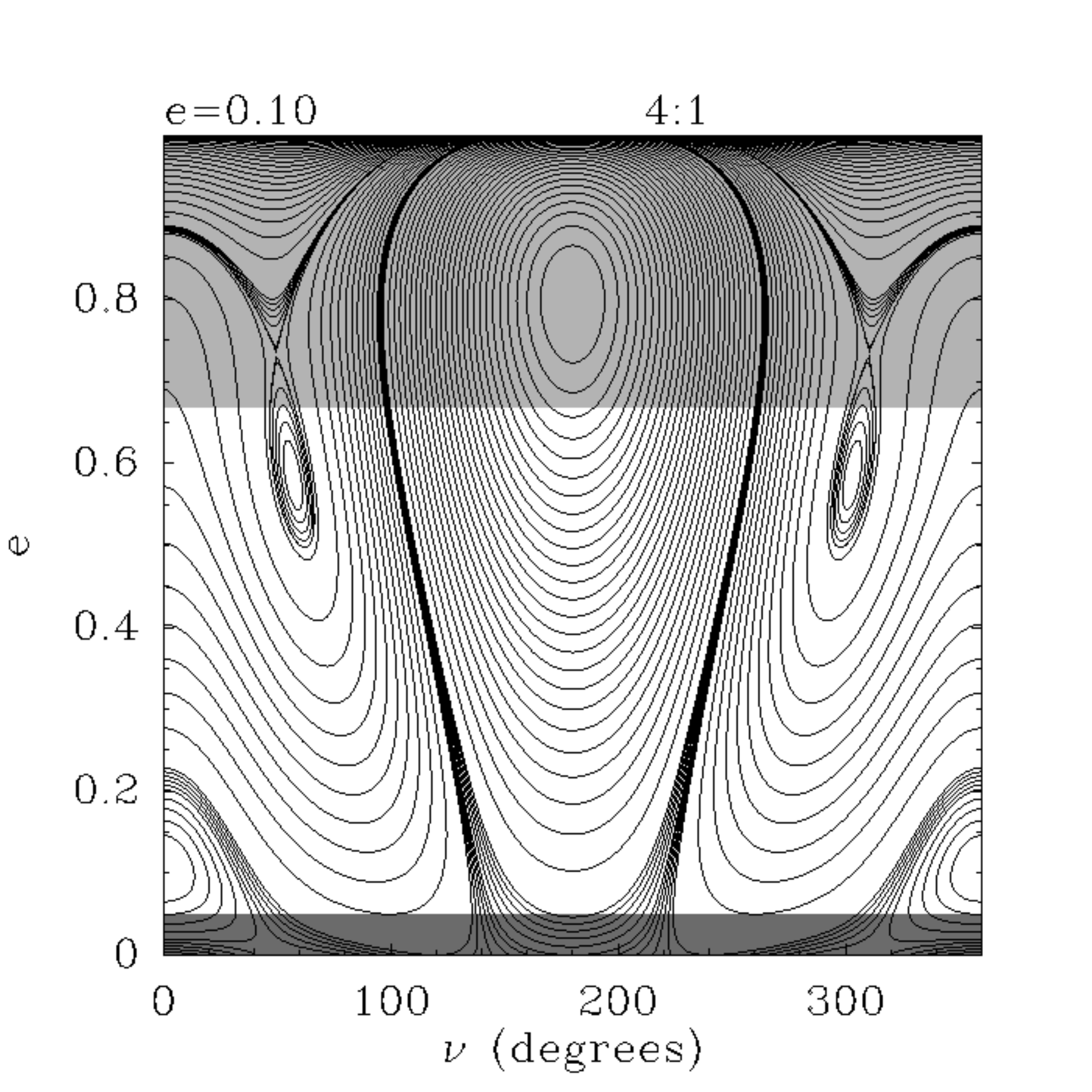}
\includegraphics[width=0.3\textwidth,height=0.3\textwidth]{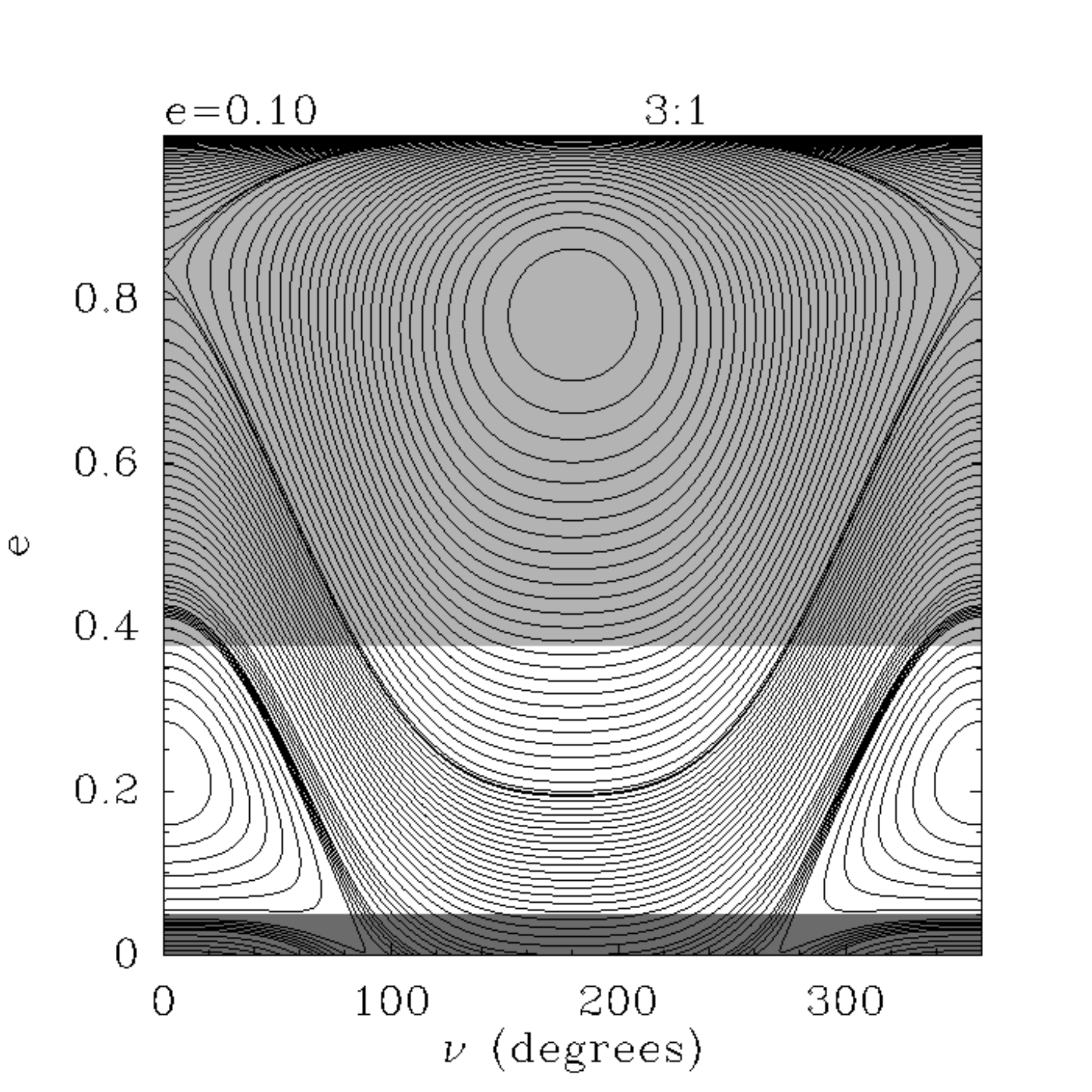}
\includegraphics[width=0.3\textwidth,height=0.3\textwidth]{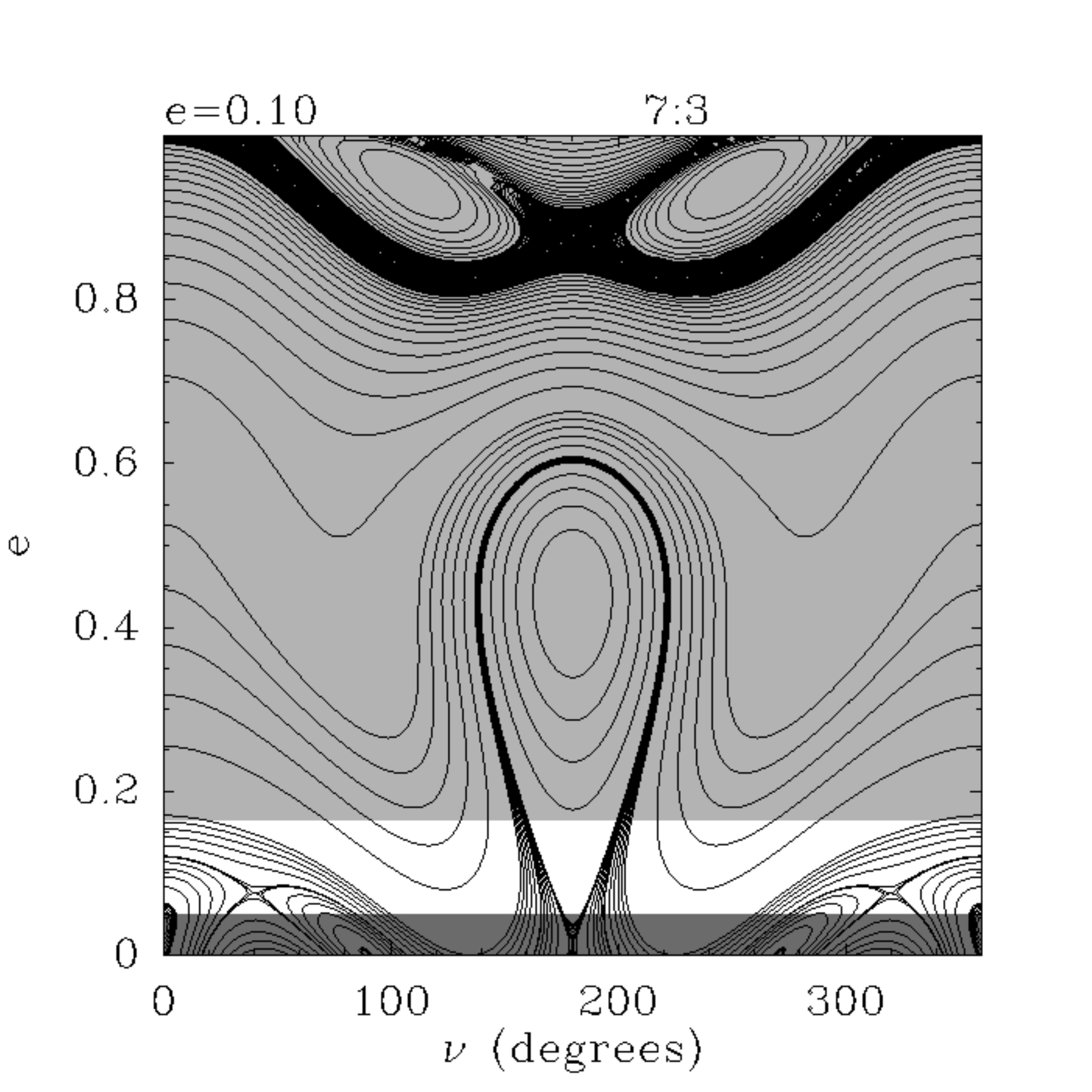}}
\makebox[\textwidth]{\includegraphics[width=0.3\textwidth,height=0.3\textwidth]{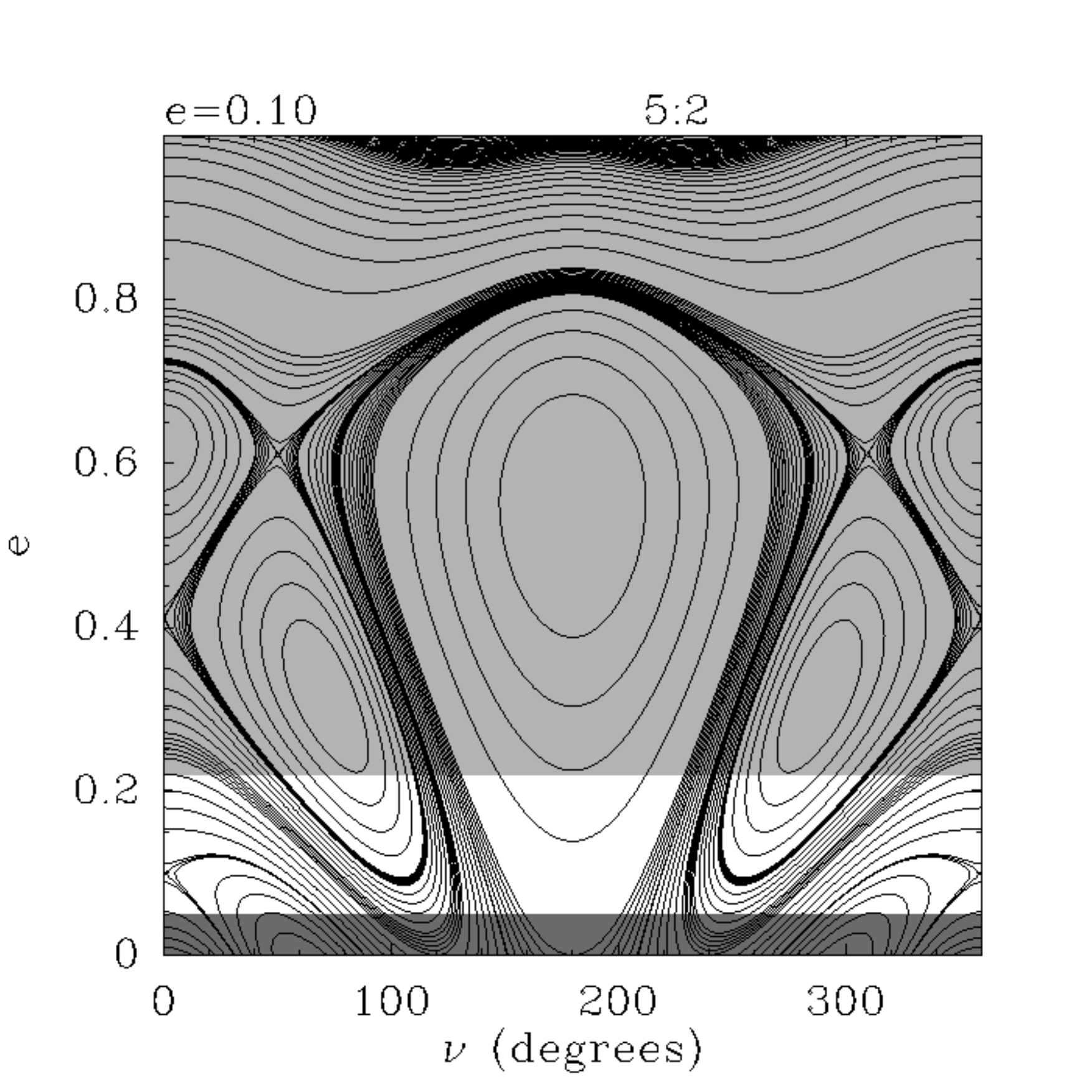}
\includegraphics[width=0.3\textwidth,height=0.3\textwidth]{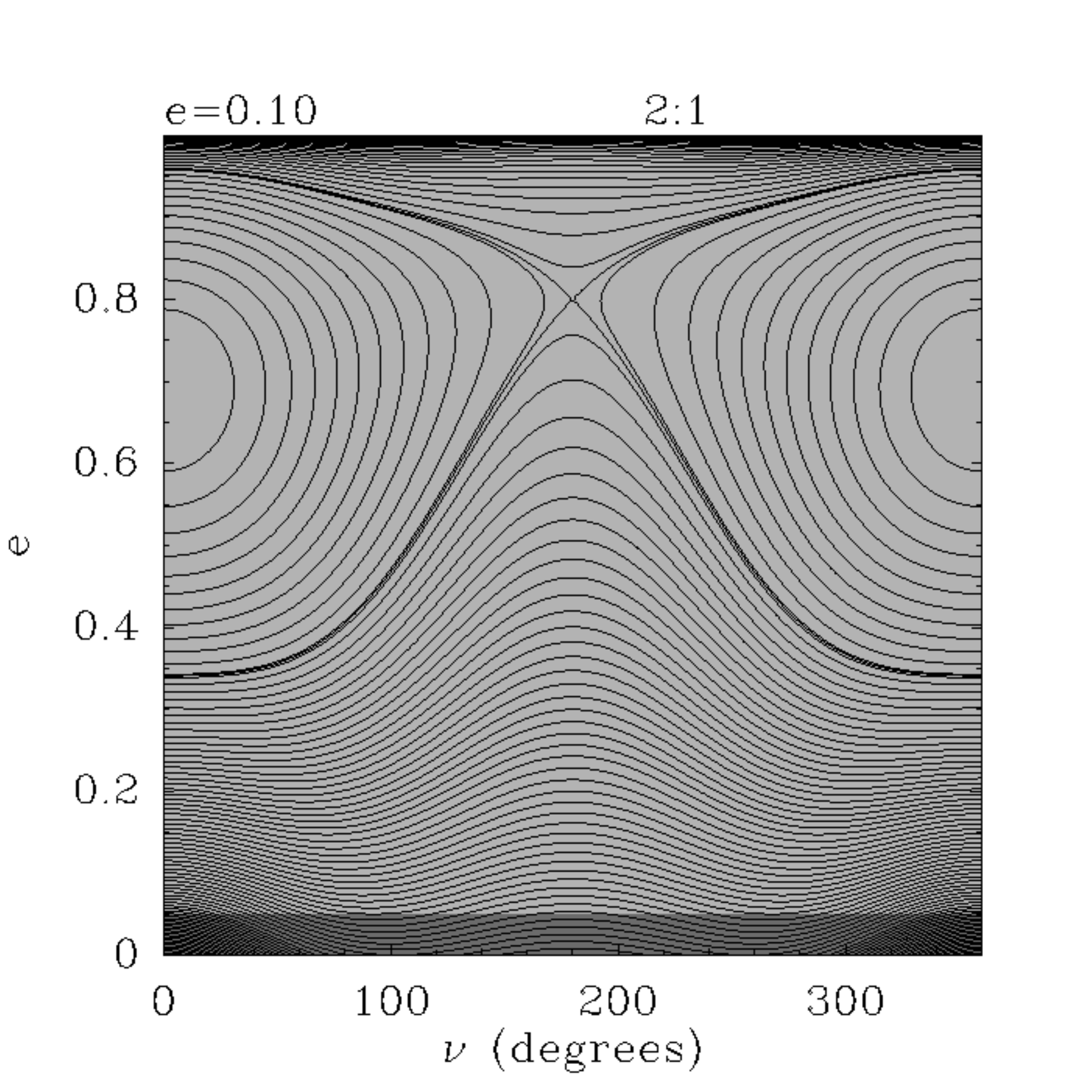}}
\caption[]{Theoretical phase diagrams of the MMRs that we investigated. Our initial conditions are figured in dark grey and the chaotic zone of Fom c in light grey. The chaotic zone of Fom c is considered to extend from $3.5 \mathrm{R_H}$ inner to the periastron of Fom c, to $3.5 \mathrm{R_H}$ outer to the apastron of Fom c. Particles which start on a trajectory allowing it to cross the chaotic zone may then be scattered and set on a Fom b-like orbit.}
\label{fig:phase_space}
\end{figure*}

Hamiltonian $H'$ has still two degrees of freedom. During the $N$-drift however, the amplitude of the libration is roughly conserved 
\citep{1993CeMDA..57...99M,1995Icar..115...60M}. If we consider orbits with 
zero libration amplitude, then the value of $\sigma$ is fixed, and so the 
value of the semi-major axis $a$. $H'$ can then be time-averaged over the last remaining fast variable, that is, $\lambda_c$, which causes $H'$ to reduce to one degree of freedom, depending on variables $(\nu,e)$ only. Constant level curves of this simplified Hamiltonian may be drawn in $(\nu,e)$ space to build a phase-space diagram. $H'$ is numerically evaluated for a series of  $(\nu,e)$ in a grid and level curves are drawn. The result is shown in 
Fig.~\ref{fig:phase_space}. Numerical experiments \citep[See Figure 13 of][]{2000Icar..143..170B} showed that real trajectories are still accurately described by these phase-space diagrams as long as the libration amplitude of resonant particles remain moderate. 

The averaged $H'$ depends actually on $\nu$, $e$, $e_c$ and $m_c$. But as $a$ is constant (its value is fixed by the MMR condition), all variations of $H'$ are carried by the second term in Eq.~(\ref{hamil0}), that is, the disturbing function, which is proportional to $m_c$. Consequently, the topology of $H'$ is independant from $m_c$. The phase space diagrams in Fig.~\ref{fig:phase_space} can be drawn for any given MMR with fixed $e_c=0.1$ value, in  $(\nu,e)$ space.
   
For each MMR investigated, we present a phase space diagram in $(\nu,e)$ space in Fig.~\ref{fig:phase_space}. The secular evolution of Fom~b can be seen following one of the level curves of $H'$, starting in the bottom of the diagram, that is, at low eccentricity.
 We put this in perspective with the crossing of the chaotic zone of Fom c, which allows to evaluate the behaviour expected from the MMRs investigated in this paper.

It is expected that the 2:1, 7:3, 5:2, 3:1, and 4:1 MMRs allow test-particles to cross the chaotic zone of Fom c, where they will be allowed to be scattered and possibly be set on a Fom b-like orbit. This is not the case for the 8:3 and 7:2 MMRs, although in the solar system asteroid belt, they generate high eccentricity particles thanks to an overlap with a secular resonance with Saturn. Such a fortuitous configuration has only few chance to apply here. Moreover, as our simulations only consider Fom c as massiver perturber, no secular resonance is to be expected here. We nevertheless decided to keep these MMRs in our set of simulations in order to quantify this effect.

In low eccentricity regime, the radial extent of a MMR is typically $\sim 0.1$\%\ of the semi-major axis of the perturber, that is, $\sim 0.1$ AU here \citep{1996Icar..120..358B}. At high eccentricity, it is typically $\sim 20$ times wider, that is, $\sim 2\,$AU. We considered therefore that a radial extent of several AU is largely sufficient to investigate a particular resonance. Resonances were thus examined closely by distributing the semi-major axes of the test particles over 5 AU\,wide ranges centered on the theoretical location of the resonance (Table~\ref{tab:initial}).

In all cases, the system was evolved over 100 Myr, that is $\sim 1/4$ of the age of Fomalhaut, using the symplectic N-body code SWIFT-RMVS \citep{1994Icar..108...18L}, where the orbit of Fom c remained fixed, and thus, the back-reaction of test-particles on Fom c is not computed. We used a typical timestep of $\sim 1/20$ of the smallest orbital period. This ensures a conservation of energy with a typical error of $\sim 10^{-6}$ on relative energy. We took snapshots of the particles orbits every $10^{5}$ yrs. 


As explained above, the topology  of the secular interaction Hamiltonian, and hence the diagrams of Fig.~\ref{fig:phase_space}, are nearly independent from Fom c's mass. $m_c$ nevertheless directly affects the timescale of the secular evolution. As the disturbing function in Eq.~(\ref{hamil0}) is proportional to $m_c$, the speed at which the level curves are explored in Fig.~\ref{fig:phase_space},
is also proportional to $m_c$. \citep{1996Icar..120..358B} showed in the 4:1 MMR case that for a mass ratio $m_c/M=10^{-3}$, the timescale to reach high eccentricity values starting from low values is typically $\sim 10^4$ times the orbital period of the perturbing planet. Fom~c's orbital period can be estimated to $\sim 800\,$yrs. With  $m_c/M=1.5\times10^{-3}$ assumed here, it should take at most $\sim 5\,$Myr for a Fom b progenitor trapped in MMR to reach the chaotic zone starting from low eccentricity. This value must be considered as an order of magnitude estimate as it may depend on the MMR under consideration. Figure~\ref{fig:process} shows indeed the simulated evolution of a particle in 5:2 MMR which reaches the chaotic zone within a bit less than 2\,Myrs.


In the next section, we present our findings about the production of orbits similar to that of Fom b in our scenario, that is, the ability of the putative Fom c to set much less massive bodies on orbits similar to that of Fom b when these originate from the parts of the system inner to Fom c.


\section{Results}\label{sec:results}
In this section, we first retrieve general results from Run~A and then present results of Runs~B--H for individual MMRs. In our simulations, we identified particles which were set on a Fom b-like orbit, which we defined as an orbit with eccentricity and semi-major axis in the 95\% level of confidence intervals found by \citet{2014A&A...561A..43B}, that is with $\mathrm{e}\in [0.69 ; 0.98]$ and $\mathrm{a}\in [81 ; 415]\,$AU. Other constraints have been derived by \citet{2014A&A...561A..43B} regarding the orientation of the orbit of Fom b: it is almost coplanar with the outer Kuiper-belt, and more or less apsidally aligned with it. However, these constraints are weaker than those on the semi-major axis and eccentricity. Therefore, we will examinate the orientation of the Fom b-like orbits that we identified in a second time. All these constraints are summarized in Table~\ref{tab:constraints}.

\begin{table}[htbp]
\caption{Summary of the constraints on the orbit of Fom b as found by \citet{2014A&A...561A..43B}.}
\label{tab:constraints}
\begin{tabular*}{\columnwidth}{@{\excs}lll}
\hline\hline\noalign{\smallskip}
Parameter & Value & Remark \\
\noalign{\smallskip}
\hline\noalign{\smallskip}
 $\mathrm{a}\,$(AU) & 81--415 & 95\% level of confidence \\
 $\mathrm{e}$ & 0.69--0.98 & 95\% level of confidence \\
 $\mathrm{i}$ ($^{\circ}$) & 0--29  & 67\% level of confidence \\
 $\nu$ ($^{\circ}$) & $\pm$ 30--40 & $\sim70\%$ level of confidence\\
\noalign{\smallskip}\hline
\end{tabular*}
\end{table}

\subsection{Broad distribution, inner to the putative Fom c}

\begin{figure}[htbp]
\resizebox{\hsize}{!}{
\includegraphics[width=0.45\textwidth,height=0.35\textwidth]{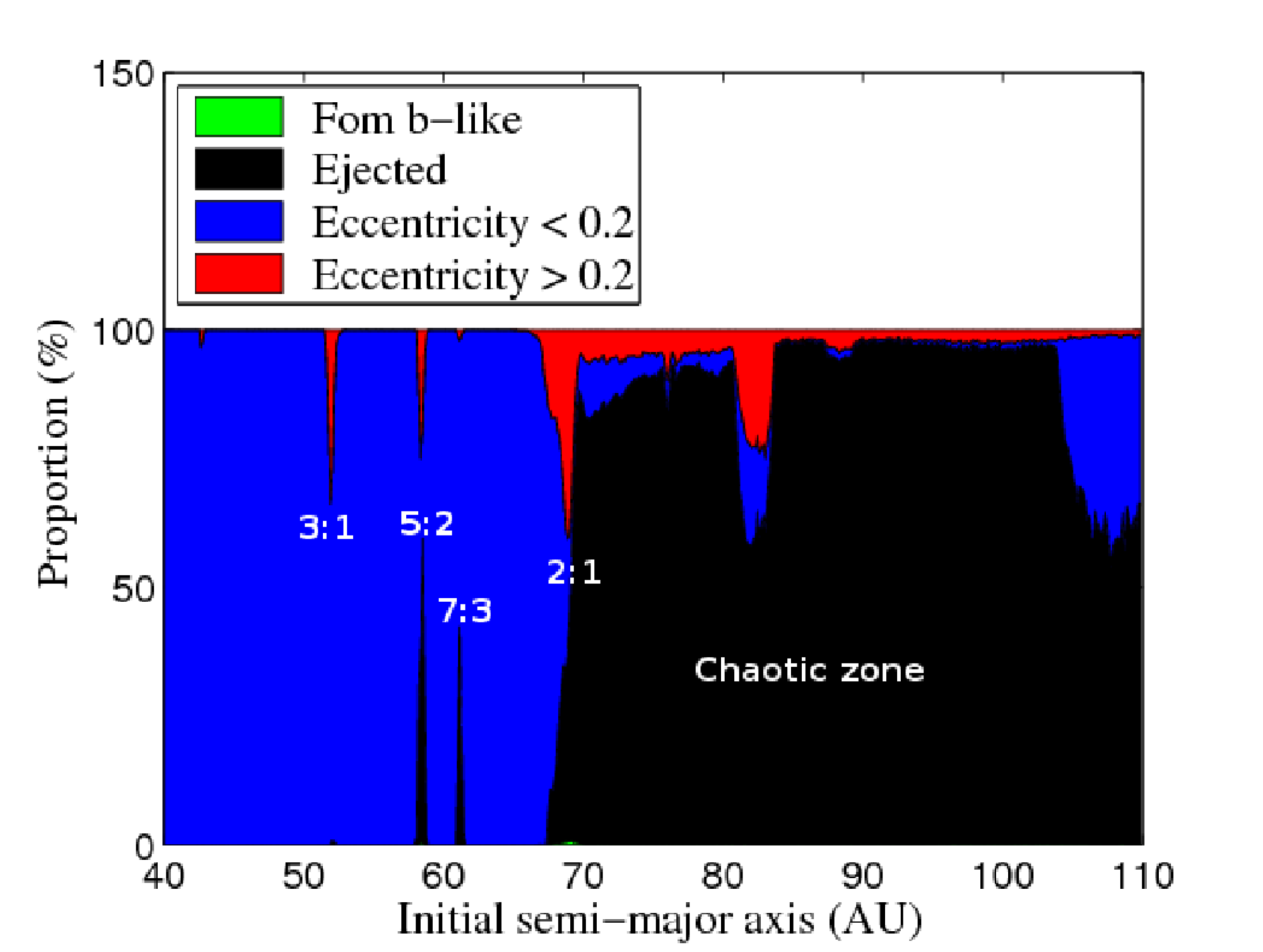}}

\resizebox{\hsize}{!}{\includegraphics[width=0.45\textwidth,height=0.35\textwidth]{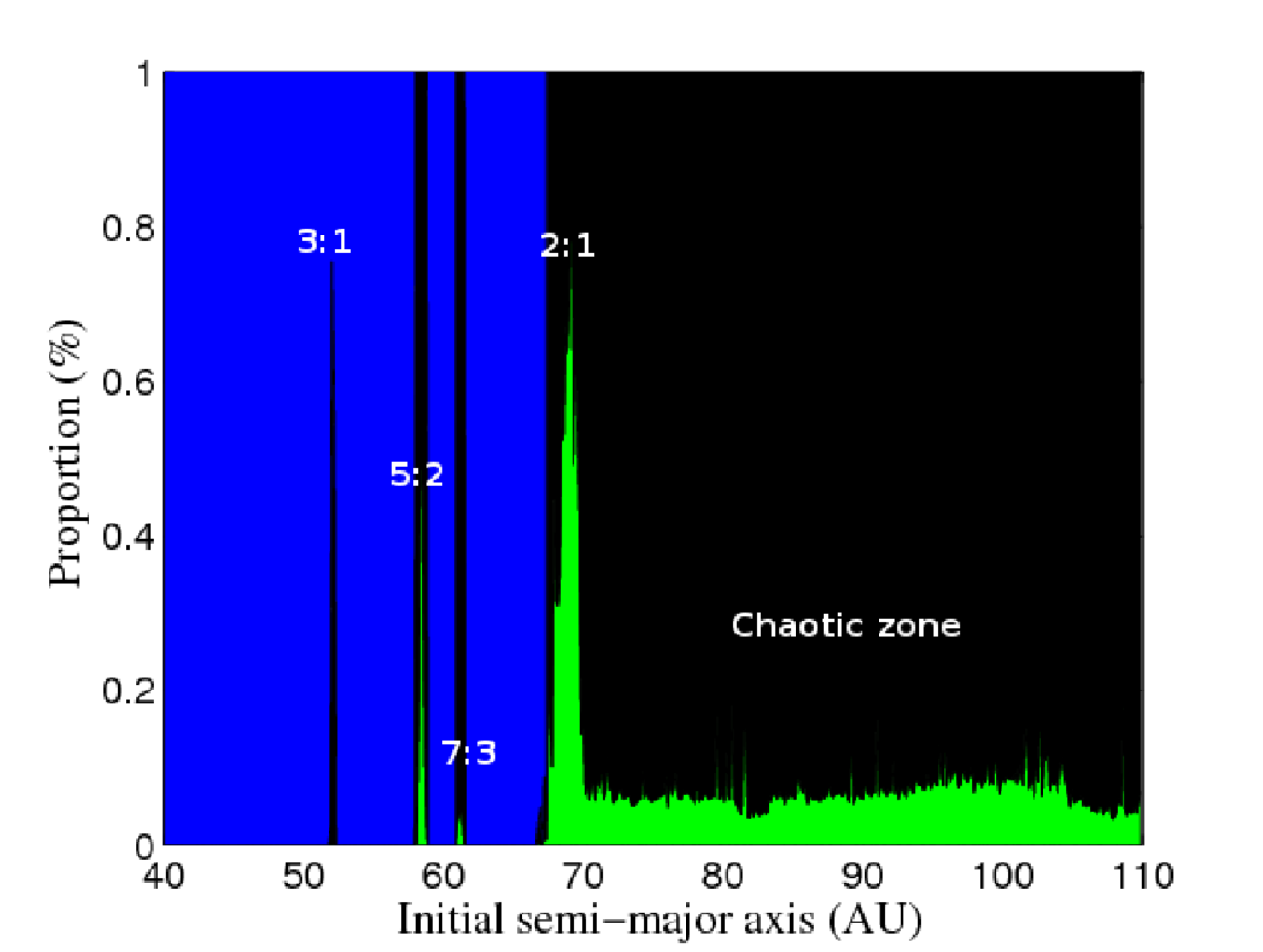}}
\caption[]{Proportion of the time spent on orbits of different dynamical status as a function of the initial semi-major axes of the test-particles. The bottom-panel is a zoom-in of the top-panel: indeed, since the total proportion of snapshots in a Fom b-like orbit dynamical status is less than $1\%$, they are not visible on the top-panel.}
\label{fig:propALL}
\end{figure}

We first investigated the dynamical status exhibited by the test-particles, integrated over the 100 Myr of the run, as a function of their initial semi-major axes in Fig.~\ref{fig:propALL}. In the top-panel, the chaotic zone of Fom c shows through a large proportion of unbound orbits above $\sim 70\,$AU, while low eccentricity orbits -- $\mathrm{e} < 0.2$ -- were preferentially adopted below this limit. The resulting width of the chaotic zone is $\sim 4.5\,\mathrm{R_H}$, and is therefore slighlty greater the theoretical one, which is $\sim 3 \mathrm{R_H}$ inner to the semi-major axis fo Fom c. Here the inner boundary rather corresponds $\sim 3 \mathrm{R_H}$ inner to periastron of Fom c. Indeed, the theoretical width of the chaotic zone we used was evaluated in the context of a circular planet, whereas Fom c is on a moderately eccentric orbit. The width of the chaotic zone created by an eccentric planet was studied by \citet{2012MNRAS.419.3074M}, and the theoretical width evaluated in this context is $\sim 4 \mathrm{R_H}$, which is more in accordance with our results.
The total proportion of snapshots in a Fom b-like orbit dynamical status is less than $1\%$, and thus they are not visible here. On the bottom-panel of Fig.~\ref{fig:propALL}, we present a zoom-in of the top-panel to show them. 

As expected, MMRs increased the eccentricity of test-particles, and possibly led them to leave the system. 
Although it was able to increase the eccentricity of test-particles, the 4:1 MMR did not to generate any Fom b-like orbit. The density of snapshots in a Fom b-like status in the cases of the 3:1 and 7:3 MMRs was low compared to that of the chaotic zone, while interestingly, this density was greater in the cases of the 5:2 and 2:1 MMRs. This does not mean that the probability for being set on a Fom b-like orbit is greater for these MMRs, but rather that Fom b-like orbits generated from these MMRs are more stable, as one can see in Table~\ref{tab:runAproba}, where we summarise the average time spent by a test-particles on a Fom b-like orbit as a function of their origin, and show for comparison the distribution of the origin of these test-particles. 
The total proportion of particles which were set at a moment or another on a Fom b-like orbit in our run is $\sim 20 \%$, where indeed $\sim 90\%$ of these test-particles originated from the chaotic zone, and the 5:2 and 2:1 MMRs produced the Fom b-like orbits with the largest mean lifetimes ($\sim 1\,$Myr), although these remain largely inferior to the age of the system.

\begin{table}[htbp]
\caption{Summary of the results of Run~A: distribution of the test-particles being set on a Fom b-like orbit and average time spent by a particle on a Fom b-like orbit, $\mathrm{\bar{t}_{Fom b}}$.}
\label{tab:runAproba}
\begin{tabular*}{\columnwidth}{@{\excs}lll}
\hline\hline\noalign{\smallskip}
Dynamical status & Distribution of Fom b-like &$\mathrm{\bar{t}_{Fom b}}$\\
 relative to Fom c &  particles (\%)  & (Myr) \\
\noalign{\smallskip}
\hline\noalign{\smallskip}
 3:1 MMR & $2.0\times 10^{-2}$ & 0.13 \\
 5:2 MMR & 1.15 & 1.0 \\
 7:3 MMR & 0.45 & 0.2 \\
 2:1 MMR & 5.35 & 1.4 \\
 Chaotic zone & 93.0 & 0.21 \\ 
\noalign{\smallskip}\hline
\end{tabular*}
\end{table}

\begin{figure}
\resizebox{\hsize}{!}{\includegraphics[width=0.7\textwidth,height=0.6\textwidth]{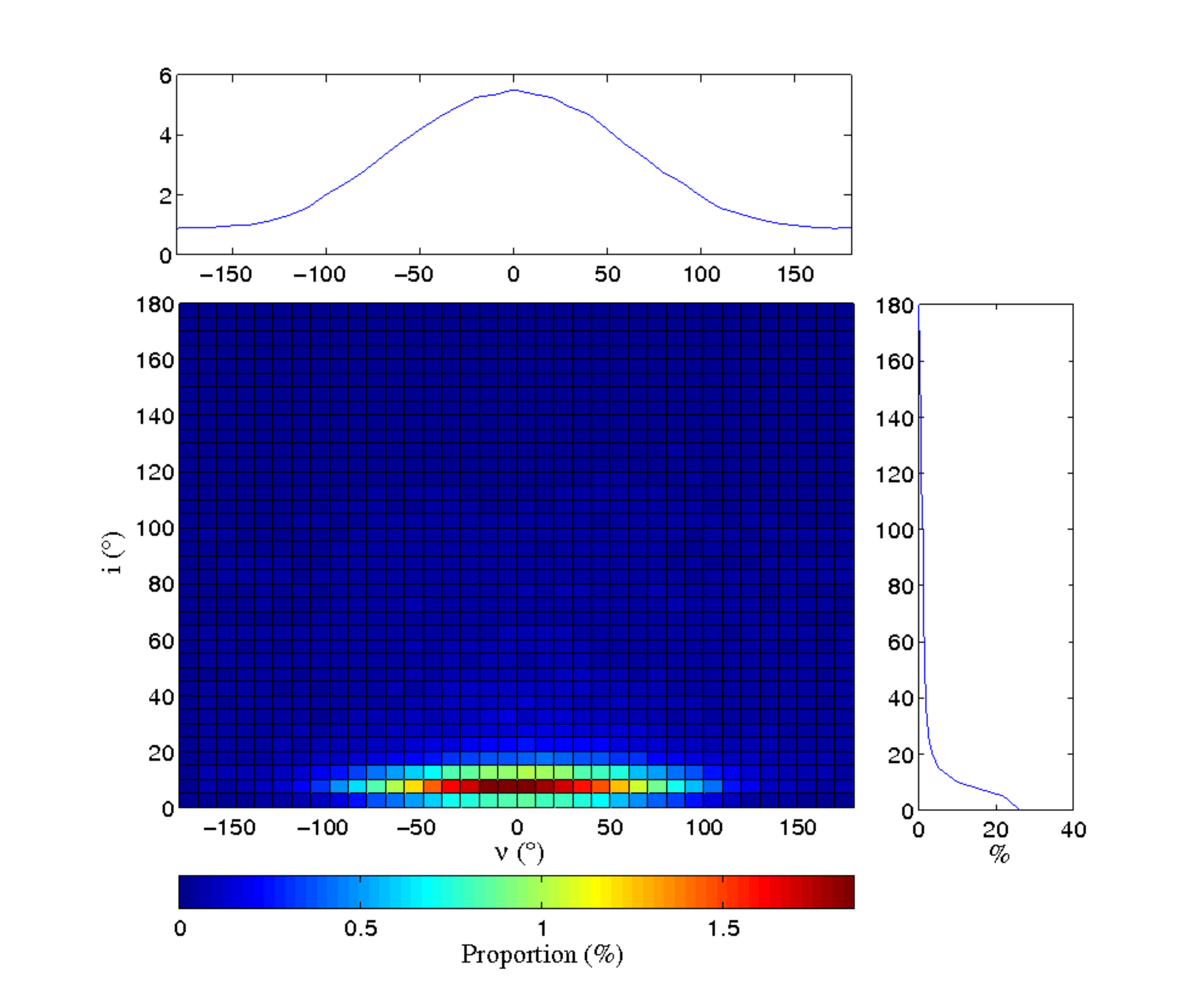}}
\caption[]{Distribution in inclination and longitude of periastron with respect to that of Fom c of Fom b-like orbits.}
\label{fig:fombALL_orient}
\end{figure}

In our definition of a Fom b-like orbit, we did not take into account the orientation of these orbits. However, constraints on the orbit of Fom b also showed that it is rather coplanar and apsidally aligned with the belt (see Table~\ref{tab:constraints}), which, in our scenario, involves that the orbit of Fom b is also coplanar and apsidally aligned with that of Fom c, since it is the planet that shapes the belt and gives it its apsidal orientation.
Therefore, we show in Fig.~\ref{fig:fombALL_orient} the orientation of Fom b-like orbits, that is their inclination $i$ and the direction of their periastron with respect of that of Fom c, $\nu$. A significant proportion of them corresponded to these criterions: indeed, $\sim 40\%$ of the Fom b-like orbits formed had $i \in [0^\circ,30^\circ]$ and $\nu \in [-40^\circ,40^\circ]$, that is, in the $\sim 70\%$ level of confidence.
This shows that the production of orbits fully comparable to that of Fom b, either in terms of semi-major axis and eccentricity, but also in terms of relative inclination to the disk and apsidal orientation, is extremely common, even in the chaotic zone. The reasons for this to happen are discussed further in Sect.~\ref{sec:apsidal}.

The sample of Fom b-like orbits from Run~A may be sufficient to retrieve first clues on the formation of these orbits, in particular, to show that MMRs may play a crucial part here, but it is probably not sufficient to fully compare the efficiency and specificities of each MMR.
Therefore we present in the following subsection the results of Runs~B--H, that is for individual MMRs.
Since these runs achieved a better sampling of MMRs, it allowed us to examine more in depth results of Run~A.

\subsection{Individual MMRs}

The results from Run~A suggested that the 4:1, 7:2, and 8:3 MMRs do not produce any Fom b-like orbit.
They also suggested that the 3:1 and 7:3 MMRs produce rare and unstable Fom b-like orbits, while the 5:2 and 2:1 MMRs tend to be very efficient in comparison, and produce Fom b-like orbits which are the most stable, although on timescales much smaller than the age of the system ($\sim 1\,$Myr).
We present hereafter the results of Runs~B--H for individual MMRs.

\subsubsection{The 7:2 and 8:3 MMRs}

No Fom b-like orbit was produced in the Runs~C and E, that is, for the 7:2 and 8:3 MMRs respectively, which is in accordance with the results of the Run~A. 

As we have seen in the previous section from the diagrams of the 7:2 and 8:3 resonances shown in Fig.~\ref{fig:phase_space}, it was obvious that test-particles initially on low-eccentricity orbits would not be able to cross the chaotic zone of Fom c, and therefore, not be able to be scattered on a Fom b-like orbit. Indeed, in both cases, none of the test-particles of the run were ever set on an eccentricity greater than 0.2. As was mentionned previously, these two MMRs need to overlap with secular resonances to trigger an increase of the eccentricity of the test-particles. They were thus expected not to be an efficient mechanism to generate Fom b-like orbits, if not unefficient at all.

\subsubsection{The 4:1, 3:1 and 7:3 MMRs}

No Fom b-like orbit was produced in the Run~B for the 4:1 MMR, in accordance with the results of the Run~A. The diagram for this MMR shows that some of our test-particles would be expected to cross the chaotic zone of Fom c, and that the production of Fom b-like orbits would be expected. So does for the 3:1 MMR, which produced very rare Fom b-like orbits in the Run~A, which is confirmed by results of Run~D. As mentioned in the previous section, it was expected that the most inner MMRs would be less efficient at producing Fom b-like orbits, since their location require a more signifant increase for test-particles to cross the chaotic zone of Fom c than MMRs located closer to Fom c.

\begin{figure*}[htbp]
\makebox[\textwidth]{\includegraphics[width=0.35\textwidth,height=0.27\textwidth]{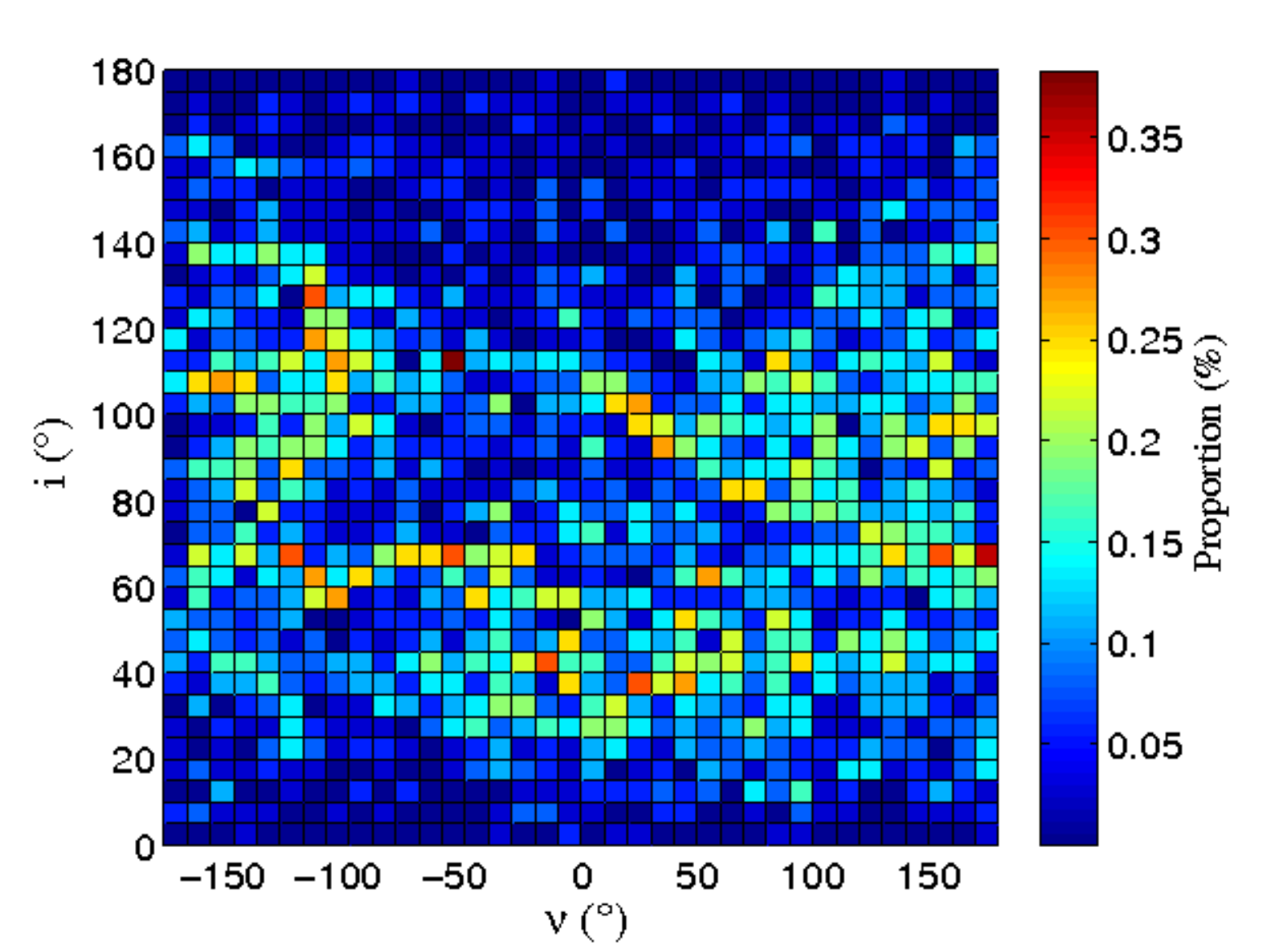}
\includegraphics[width=0.35\textwidth,height=0.27\textwidth]{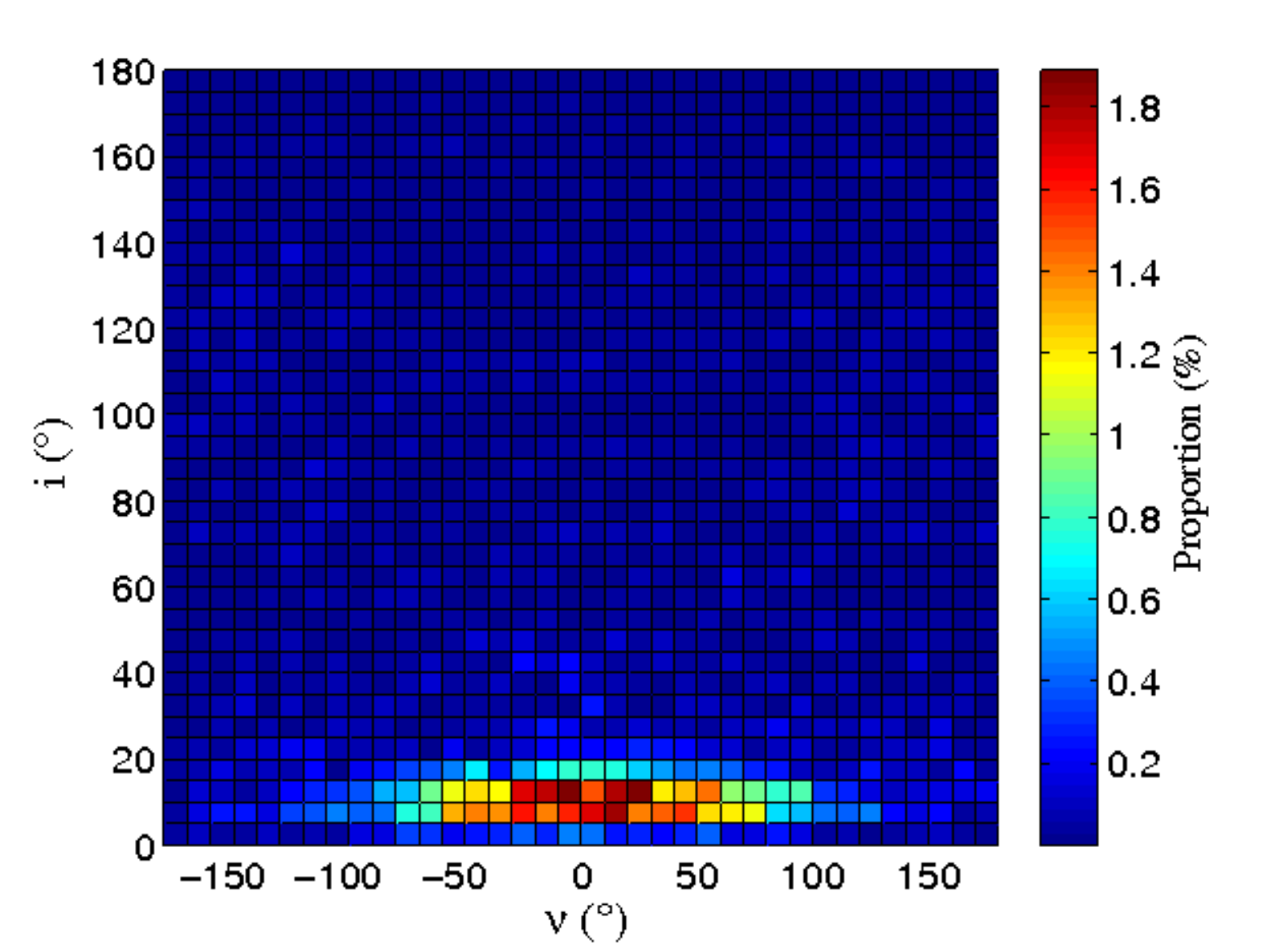}
\includegraphics[width=0.35\textwidth,height=0.27\textwidth]{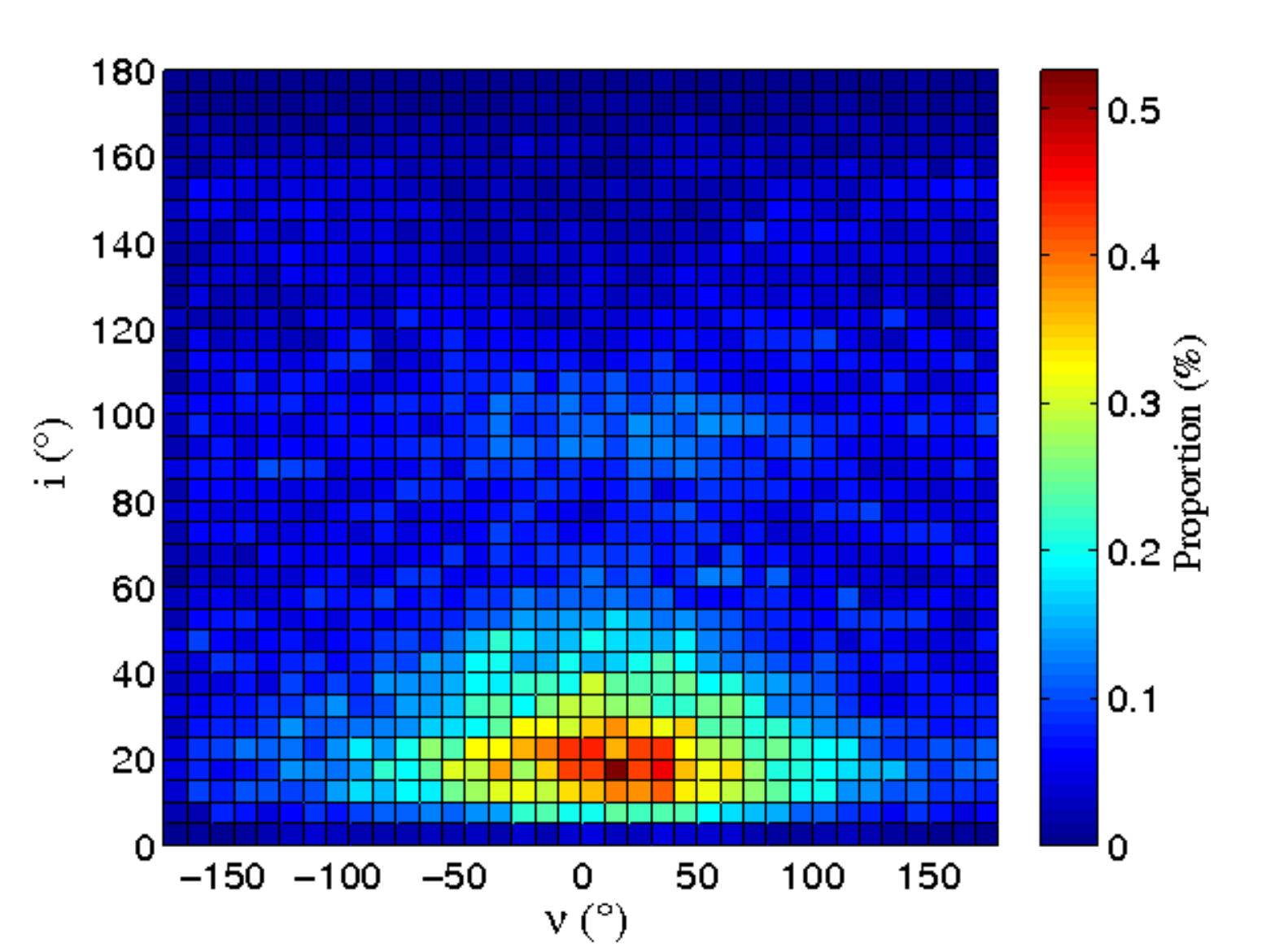}}

\makebox[\textwidth]{
\includegraphics[width=0.35\textwidth,height=0.27\textwidth]{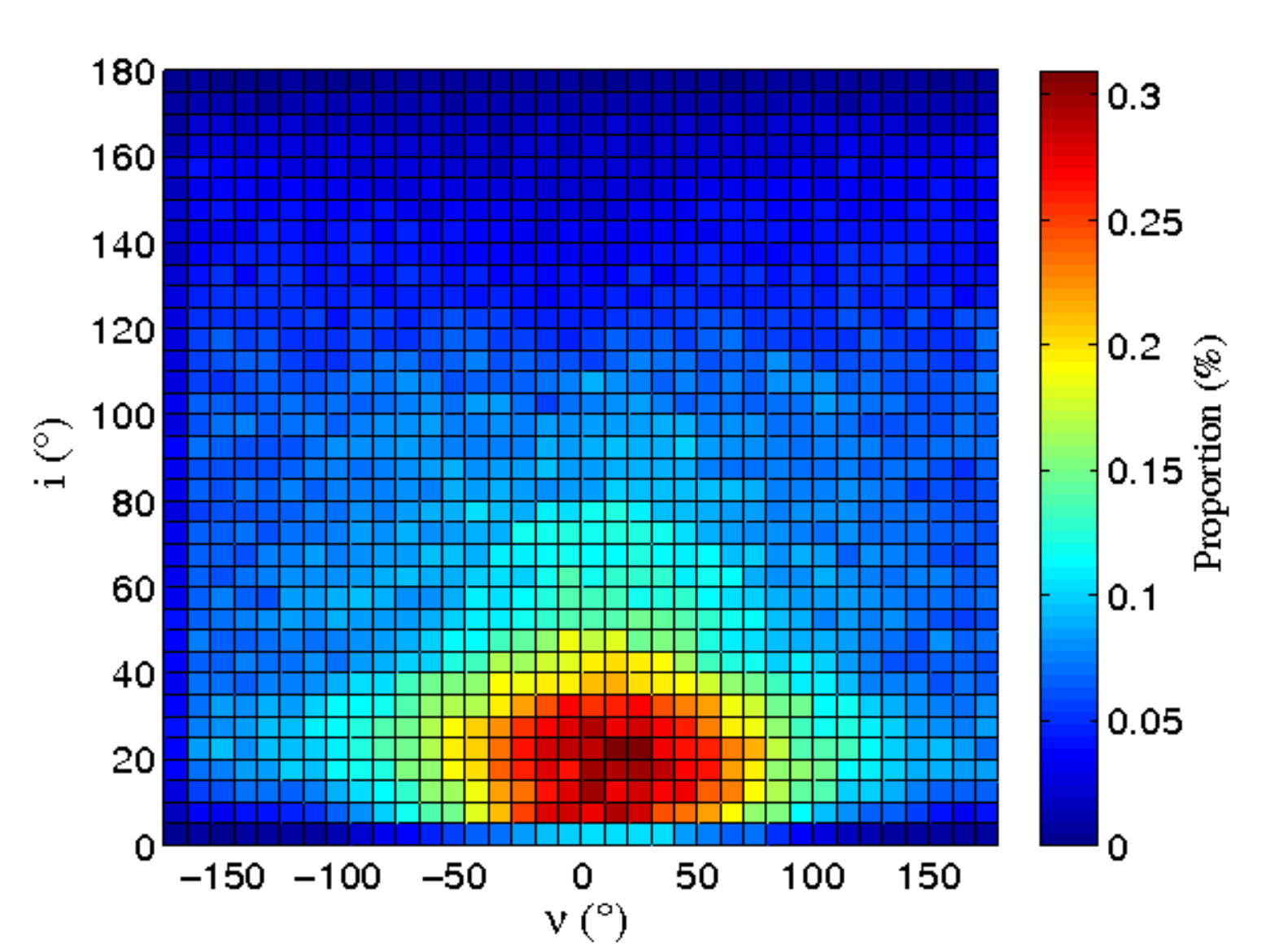}
\includegraphics[width=0.35\textwidth,height=0.27\textwidth]{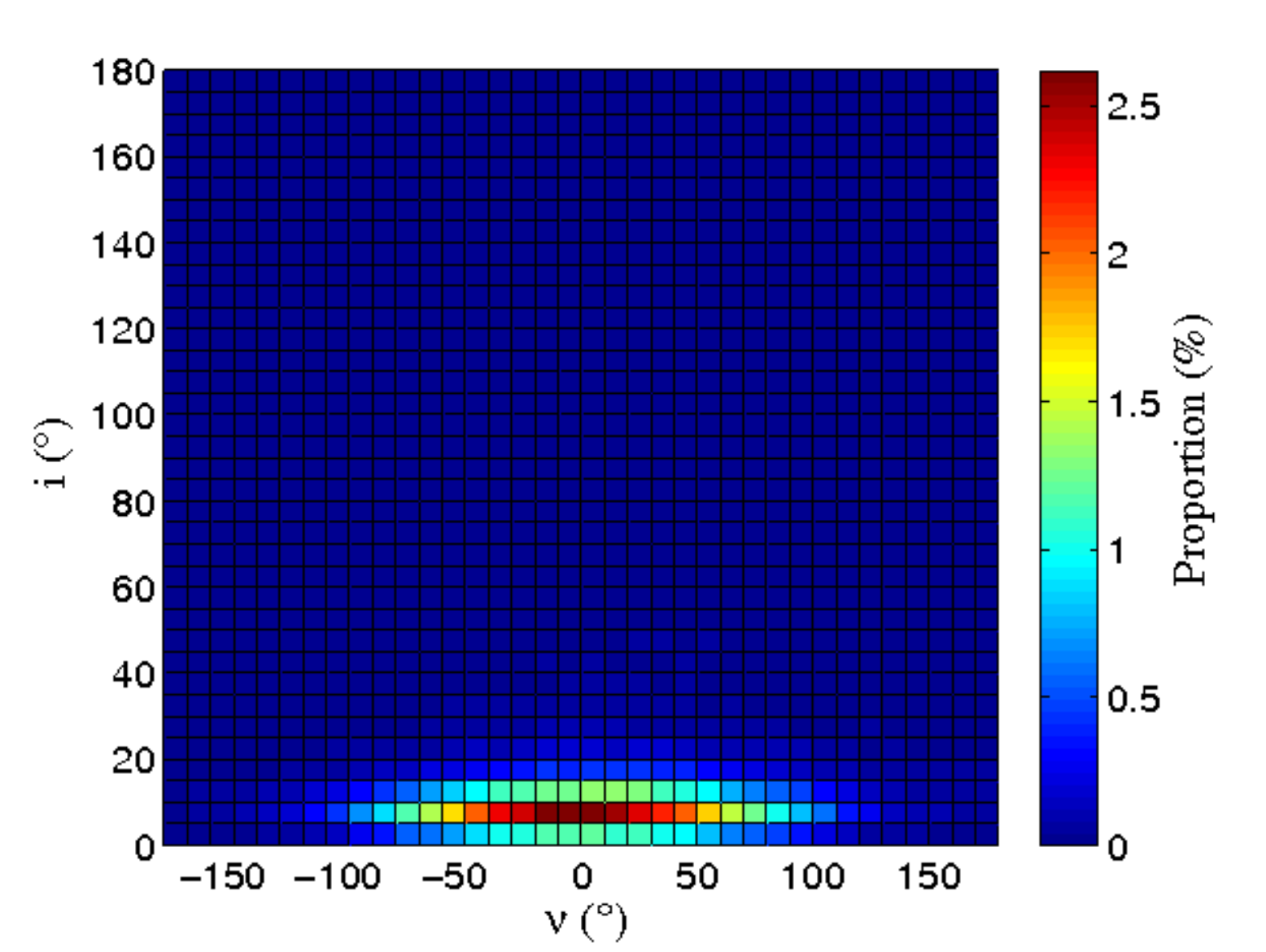}}
\caption[]{Distribution in inclination and longitude of periastron with respect to this of Fom c of Fom b-like orbits produced in the case of the 3:1, 7:2, 5:2, and 2:1 MMRs, and the chaotic zone, from top to bottom and left to right, respectively.}
\label{fig:MMR_orient}
\end{figure*}

Very interestingly, the 3:1 MMR delayed the production of Fom b-like orbits by $\sim 30-40\,$Myr compared to other MMRs, which strongly reflects the delay potentially induced by the gradual increase of the eccentricity of a resonant test-particle before it is able to be scattered. It is however  notable that the 3:1 MMR generated Fom b-like orbits with completely random orientation, which critically reduces the chance of apsidal alignement between Fom b and the dust belt (see top-left-panel of Fig.~\ref{fig:MMR_orient}).

The results of the Run~G, that is, for the 7:3 MMR, are in accordance with the results of the Run~A: the Fom b-like orbits produced were rare $(\sim 1\%)$. Moreover, they were theoretically expected as shown in the phase-diagram for this MMR. The average time spent by test-particles on a Fom b-like orbit is $\sim 0.3\,$Myr, and the maximum time for an individual particle does not exceed 40 Myr. The delay induced in the generation of Fom b-like orbits by this MMR is much smaller than the age of the system ($\sim 3\,$Myr). However, the Fom b-like orbits produced by this MMR are also similar with the observed orbit of Fom b in terms of orientation (see top-middle-panel of Fig.~\ref{fig:MMR_orient}). 

\subsubsection{The 5:2 and 2:1 MMRs}

The MMRs for which test-particles have greater probabilities to be set on a Fom b-like orbit, that is, the 5:2 and 2:1 MMRs, also produced a significant proportion of orbits with orientation comparable to that of Fom b (see top-right and bottom-left panels of Fig.~\ref{fig:MMR_orient}).

Their phase-space diagrams revealed that a great number of particles will cross the chaotic zone of Fom c (see Fig.~\ref{fig:phase_space}). In particular, the 2:1 MMR is not expected itself to sustain large increases in eccentricity, but it appears to be fortuitously located at the boundary of the chaotic zone of Fom c with our parameters, and more particularly, with the mass chosen for Fom c. The impact of the mass of Fom c will be discussed in Sect.~\ref{sec:mass_impact}. In the present case, the increase in eccentricity needed is very small and test-particles have easily crossed the chaotic zone of Fom c.
These two MMRs seem to be valid routes to form orbits comparable to this of Fom b.
However, in both cases, although rare test-particles adopted a Fom b-like orbit over more than 40 Myr, the average time spent on a Fom b-like orbit is $\lesssim 2$ Myr, and the delay induced in the generation of Fom b-like orbits by these MMRs is much smaller than the age of the system ($\sim 1-2\,$Myr)

\subsection{Summary}

Combining the results for individual MMRs and for the chaotic zone, we summarise in Table~\ref{tab:probaMMRs} different probabilities which characterize the production of Fom b-like orbits: $\mathrm{P_{Fomb}}$ is the probability to be set on a Fom b-like orbit, $\mathrm{P_{orient}}$ is the probability for a Fom b-like orbit to have an orientation compatible with that observed for Fom b, that is $i \in [0^\circ,30^\circ]$ and $\nu \in [-40^\circ,40^\circ]$, $\mathrm{P_{>10\,Myr}}$ is the probability for a Fom b-like orbit to have a lifetime greater than 10 Myr. We summarise as well the average time $\mathrm{\bar{t}_{Fom b}}$ spent by test-particles which were set on a Fom b-like orbit, and the delay in the generation of Fom b-like orbits induced by MMRs.

\begin{table}[htbp]
\caption{For each individual run that produced Fom b-like orbits, probability $\mathrm{P_{Fomb}}$ for being set on a Fom b-like orbit, that is, the proportion of the 100,000 test-particles of our initial sample set at least once on a Fom b-like orbit, average time $\mathrm{\bar{t}_{Fom b}}$ spent by these test-particles in this configuration, probability $\mathrm{P_{>10\,Myr}}$ for a Fom b-like orbit to have a lifetime greater than 10 Myr and probability $\mathrm{P_{orient}}$ for a Fom b-like orbit to have an orientation comparable to that of Fom b. We indicate as well any delay in the generation of Fom b-like orbits.}
\label{tab:probaMMRs}
\begin{tabular*}{\columnwidth}{@{\excs}llllll}
\hline\hline\noalign{\smallskip}
Dynamical status & $\mathrm{P_{Fomb}}$ & $\mathrm{\bar{t}_{Fom b}}$ & $\mathrm{P_{>10\,Myr}}$ & $\mathrm{P_{orient}}$ & Delay \\
 relative to Fom c& (\%) & (Myr) & (\%) & (\%)&  (Myr) \\
\noalign{\smallskip}
\hline\noalign{\smallskip}
 3:1 MMR & $9.7\times 10^{-2}$ & 3.8 & 11.3 & 4.4 & $\sim 30$ \\
 5:2 MMR & $3.8$ & 1.2 & 2.4 & 17.6 & $\sim 2$ \\
 7:3 MMR & $1.3$ & 0.24 & $2.2\times 10^{-1}$ & 39.6 & $\sim 3$ \\
 2:1 MMR & $20.1$ & 1.6 & 3.5 & 15.2 & $\sim 1$ \\
 Chaotic zone & $35.5$ & 0.21 & $8.5\times 10^{-2}$ & 48.5 & 0 \\ 
\noalign{\smallskip}\hline
\end{tabular*}
\end{table}

\begin{figure*}[htbp]
\makebox[\textwidth]{\includegraphics[width=0.5\textwidth,height=0.35\textwidth]{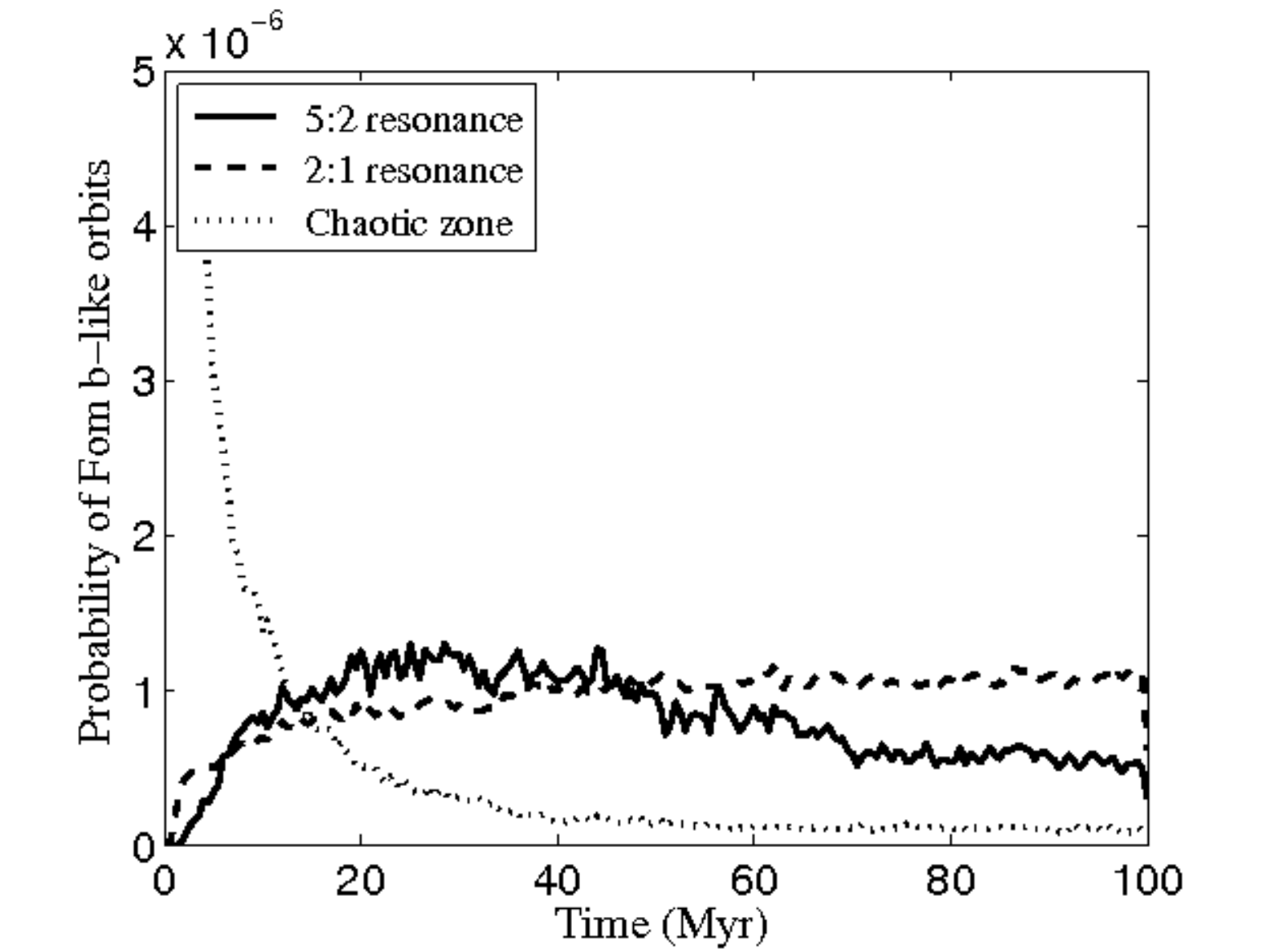}
\includegraphics[width=0.5\textwidth,height=0.35\textwidth]{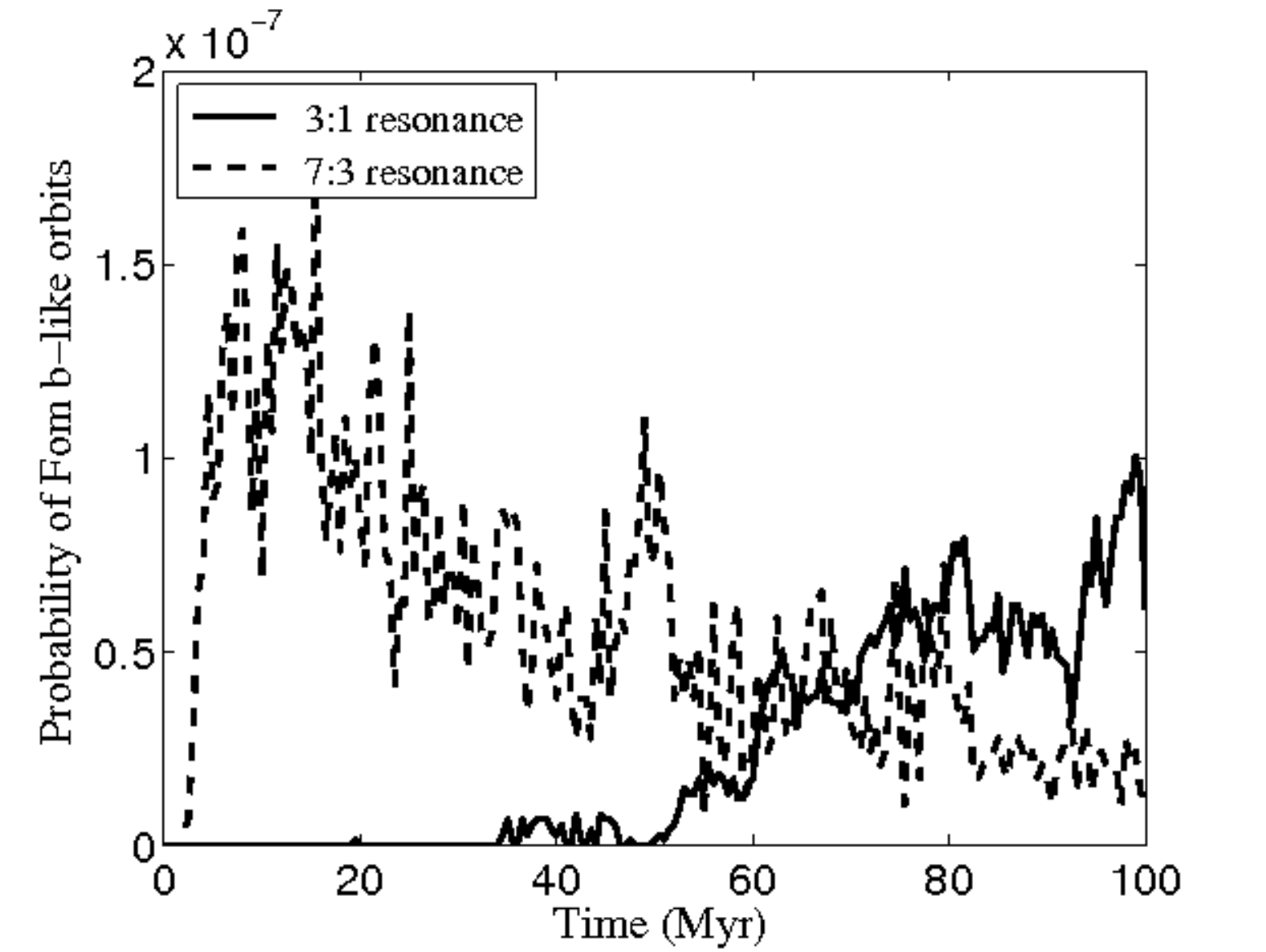}}
\caption[]{Probability to be set on a Fom b-like orbit with a lifetime greater than 10 Myr and with an orientation comparable to that of Fom b as a function of the time and the origin of the test-particles.}
\label{fig:proba}
\end{figure*}

Moreover, for each region of interest, one can retrieve the probability to be set on a Fom b-like orbit as a function of time, which, ponderated by the corresponding probabilities $\mathrm{P_{Fomb}}$, $\mathrm{P_{>10\,Myr}}$ and $\mathrm{P_{orient}}$ allowed us to fully compare the efficiency of each region of interest to produce orbits fully comparable to the orbit of Fom b and which have lifetimes greater than 10 Myr, as shown in Fig.~\ref{fig:proba}.

All these results give first insights on the most probable origin and dynamical history of Fom b in our scenario. The probability to be set on a Fom b-like orbit fully comparable to that of Fom b and which survives longer than 10 Myr is smaller by an order of magnitude for the 3:1 and 7:3 MMRs compared to the chaotic zone and the 5:2 and 2:1 MMRs. The chaotic zone of Fom c is very efficient at producing Fom b-like orbits, but these are highly unstable and the probability to be set on a Fom b-like orbit decreases very quickly during the first $\sim  10\,$Myr, where its efficiency to produce Fom b-like orbits becomes smaller than these of the 2:1 and 5:2 MMRs. Therefore, the chaotic zone, the 3:1, and the 7:3 MMRs may not be the best ways to explain the orbit and dynamical history of Fom b.

On the other hand, the 2:1 MMR is very efficient at producing Fom b-like orbits, and which have a longer lifetime. However, their orientation is much less probable to be comparable with this of Fom b. Finally, although the 5:2 MMR produces less Fom b-like orbits than the 2:1 MMR, it produces Fom b-like orbits with comparable lifetime, and additionally, produces Fom b-like orbits with orientation fully comparable to this of Fom b in a very significant proportion.
Therefore, the 5:2 and 2:1 MMRs appear to be the most probable origin of Fom b in our scenario.

Surprisingly, Fom b-like orbits originating from the chaotic zone have an orientation comparable to this of Fom b, in even greater proportions than MMRs. This is indeed surprising because one would rather have expected some specific MMRs to be able to generate such a significant tendency for apsidal alignement, since they may cause a preferential geometry of close-encounters, while random encounters in the chaotic zone would have had expected to generate randomly orientated Fom b-like orbits. Instead, the apsidal alignement feature appears to be very common, excepted for the 3:1 MMR.

We discuss our results and investigate the influence of the eccentricity and mass of Fom c in the next section. These parameters reveal to be crucial since the former controls the ability of Fom c to generate Fom b-like orbits via MMRs and the latter controls the delay in the generation these orbits.
We further investigate the mechanism that generates Fom b-like orbits and the origin of the observed common apsidal alignement in more details.


\section{Discussion}\label{sec:discussion}
In this section, we further investigate the influence of the orbital
eccentricity and mass of Fom c. We focus especially on the 2:1 and 5:2
MMRs, which, as we have seen in the previous section, seem so far to
be the best routes to have led Fom b on its current orbit in our
scenario. We also investigate further the process that generates Fom
b-like orbits and the tendency for those to be apsidally aligned with
the putative Fom c and the outer belt in a very general manner.

\subsection{Eccentricity of Fom c}

\begin{figure*}[htbp]
\makebox[\textwidth]{
\includegraphics[width=0.3\textwidth,height=0.3\textwidth]{52fom_01}
\includegraphics[width=0.3\textwidth,height=0.3\textwidth]{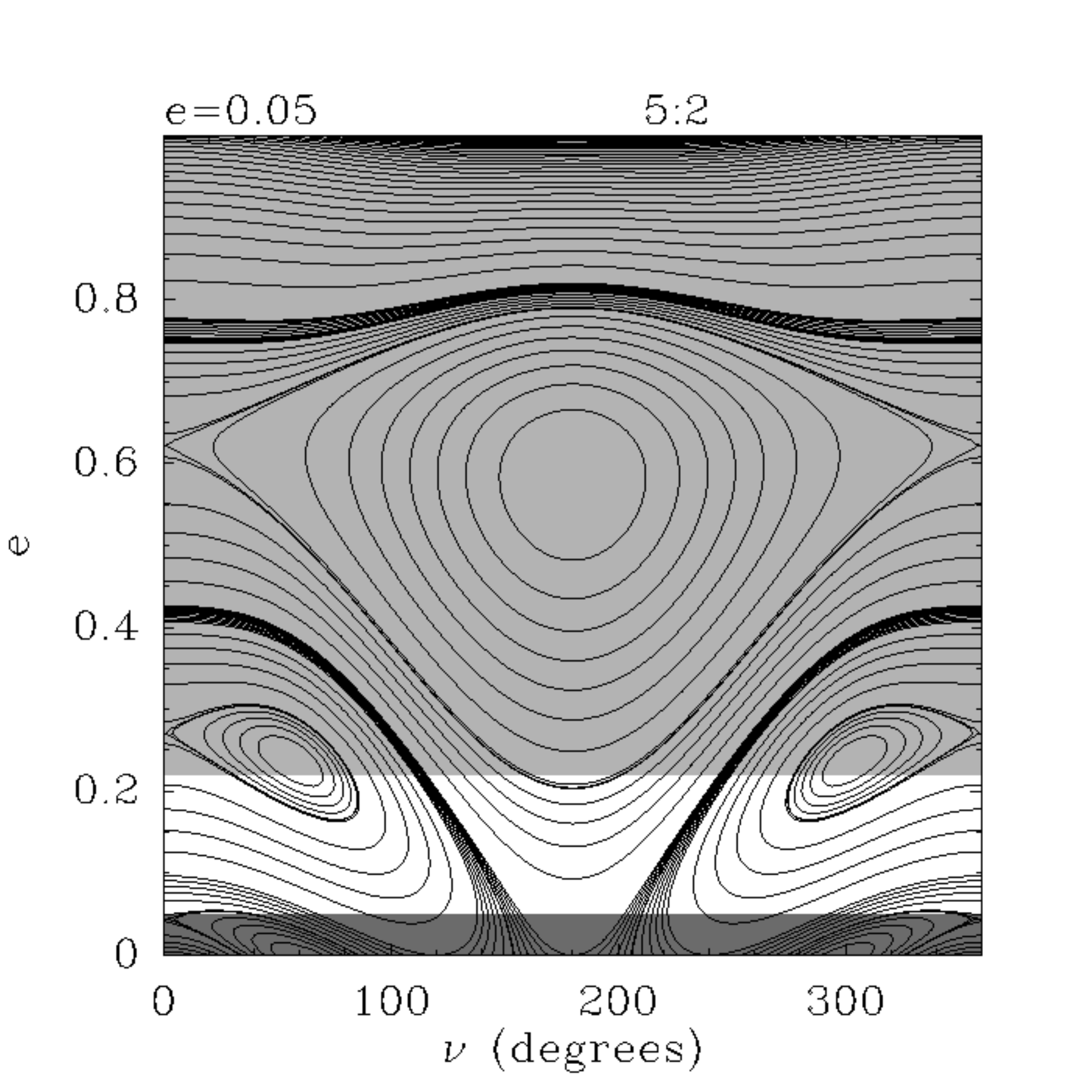}
\includegraphics[width=0.275\textwidth,height=0.275\textwidth]{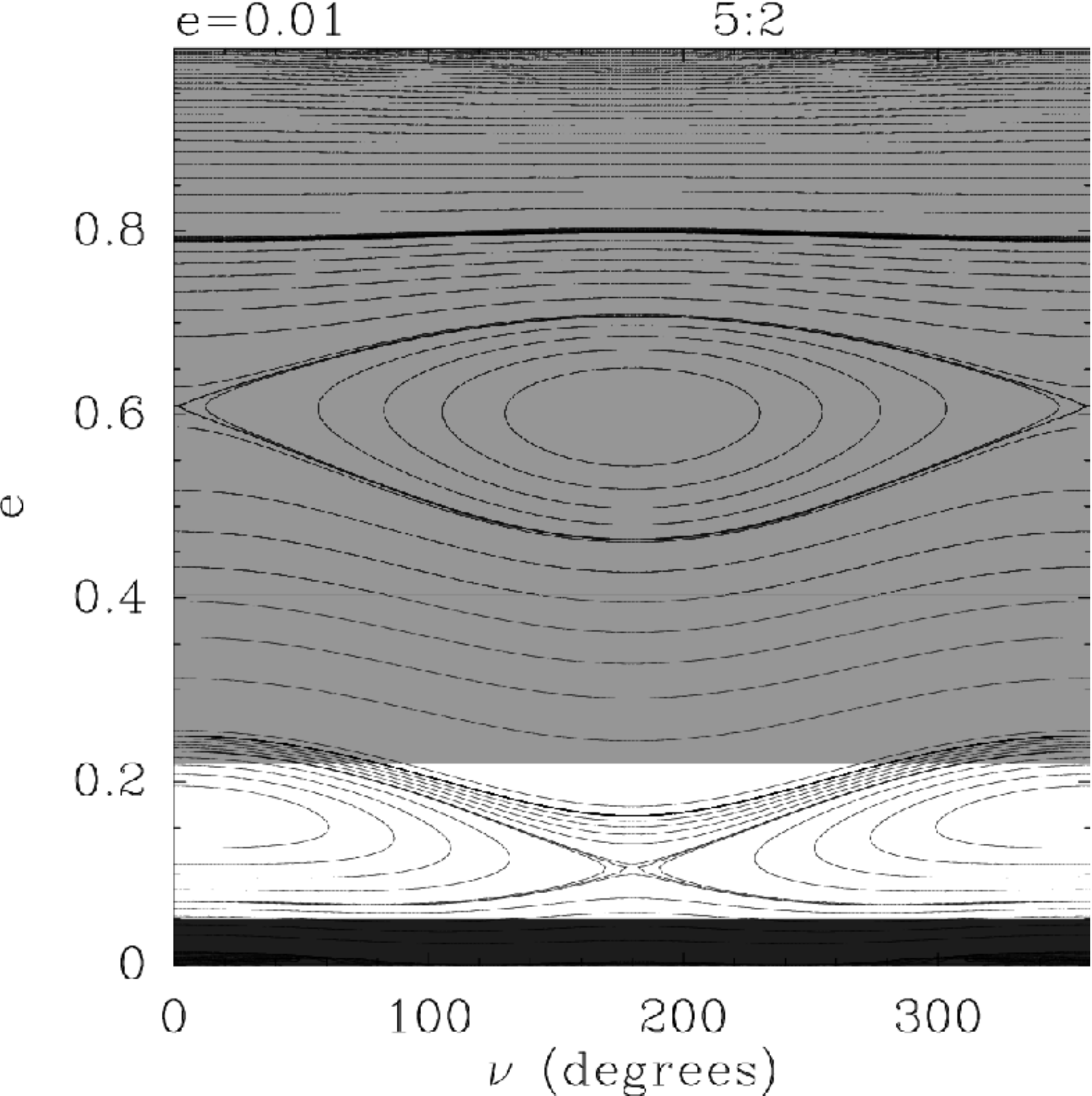}}
\caption[]{Phase diagram for the 5:2 MMR, with Fom c of eccentricity
  0.1, 0.05, and 0.01, from left to right, respectively. Our initial
  conditions are figured in dark grey and the chaotic zone of Fom c in
  light grey. The chaotic zone of Fom c is considered to extend from
  $3.5 \mathrm{R_H}$ inner to the periastron of Fom c, to $3.5
  \mathrm{R_H}$ outer to the apastron of Fom c. Particles which start
  on a trajectory allowing it to cross the chaotic zone may then be
  scattered and set on a Fom b-like orbit.}
\label{fig:MMR52_ecc}
\end{figure*}

The initial eccentricity of Fom c is a crucial paramater. Indeed, the
trajectories offered in the phase-space to test-particles in resonance
are very sensitive to the eccentricity of the perturber that creates
these resonances. An orbital eccentricity as small as 0.1 for Fom c is
actually necessary to produce Fom b-like orbits because particles are
allowed in this case to reach the eccentricities necessary for them to
cross the close encounter zone of Fom c. This can be seen on
Fig.~\ref{fig:MMR52_ecc}. Another simulation (not shown here) studying
the 5:2 resonance with a Fom c on an orbit with eccentricity 0.05
revealed as expected that the number of particles set on a Fom b-like
orbit decreases dramatically. Only $\sim 0.5\%$ of the particles of
the run were indeed set on such an orbit. Moreover, the time spent by
these on a Fom b-like orbit did not even exceed 10 Myr, with an
average of $\sim 0.3\,$Myr. Another run with a Fom c on a circular
orbit did not produce any Fom b-like orbit, as expected. Therefore,
Fom b-like orbits can be considered a natural consequence of Fom c
having an eccentricity of 0.1. Note that is is fully
  compatible with the measured eccentricity of the dust belt. If we
  believe that Fom c controls the dynamics of the dust belt, then
  secular (pericenter glow) theory, developped by \citet{1999ApJ...527..918W}, shows that the disk is
  expected to achieve a bulk eccentricity comparable to that of the
  perturbing planet.

\subsection{Mass of Fom c}\label{sec:mass_impact}

The mass of Fom c is also a crucial parameter, which controls the
ability of a given MMR to produce Fom b-like orbits by varying the
size of the chaotic zone, but also very importantly, controls
dynamical timescales, that is, the delay induced by a given MMR in the
production of Fom b-like orbits and the survival timescale of Fom
b-like orbits.

Since the Hill radius and thus the width of the chaotic zone increases
with the mass of Fom c $(\propto m_\mathrm{c}^{1/3})$, for a same
planetary semi-major axis, a less massive Fom c is expected to
generate a thinner chaotic zone that small bodies in MMR will be then
less probable to cross. Therefore, one should expect less Fom b-like
orbits to be generated with less massive Fom c.  This can be shown by
examining the minimum eccentricity needed for a test-particle in MMR
to reach the chaotic zone of Fom c, because it depends only on the
mass of Fom c for each MMR.  Indeed, if we assume that the inner boundary
of the chaotic zone of Fom c is $\mathrm{a_{in}=a_c-3R_H}$, then
Eq.~(\ref{eq:rhill}) gives:

\begin{equation}\label{eq:cond1}
a_\mathrm{in}=a_\mathrm{c}\left[1-3\left(\frac{m_\mathrm{c}}{3M_{\star}} \right)^{1/3}\right] \qquad.
\end{equation}

A particle will cross the chaotic zone as soon as it apoastron reaches the inner boundary of the chaotic zone. The apoastron $Q$ of a test-particle in MMR reads:

\begin{equation}\label{eq:apoastron}
Q=a_\mathrm{MMR}(1+e) \qquad,
\end{equation}

where $e$ is the eccentricity of the test-particle 
and $a_\mathrm{MMR}$ is the semi-major axis of the resonance. 
If we assume the particle to be trapped in a $p+q:p$ MMR
($p$ and $q$ integers) with Fom c, then we have:

\begin{equation}\label{eq:MMRloc}
a_\mathrm{MMR}= a_\mathrm{c} \left(\frac{p}{p+q} \right)^{2/3}\qquad.
\end{equation}

Consequently, the particle crosses the choatic zone only if 
$e\ge e_\mathrm{min}$, where $e_\mathrm{min}$ reads
\begin{equation}\label{eq:cond3}
e_\mathrm{min}=\left[1-3\left(\frac{m_\mathrm{c}}{3M_{\star}} \right)^{1/3}\right]
\left(\frac{p}{p+q} \right)^{-2/3}-1 \qquad.
\end{equation}

\begin{figure*}[htbp]
\makebox[\textwidth]{\includegraphics[width=0.3\textwidth,height=0.3\textwidth]{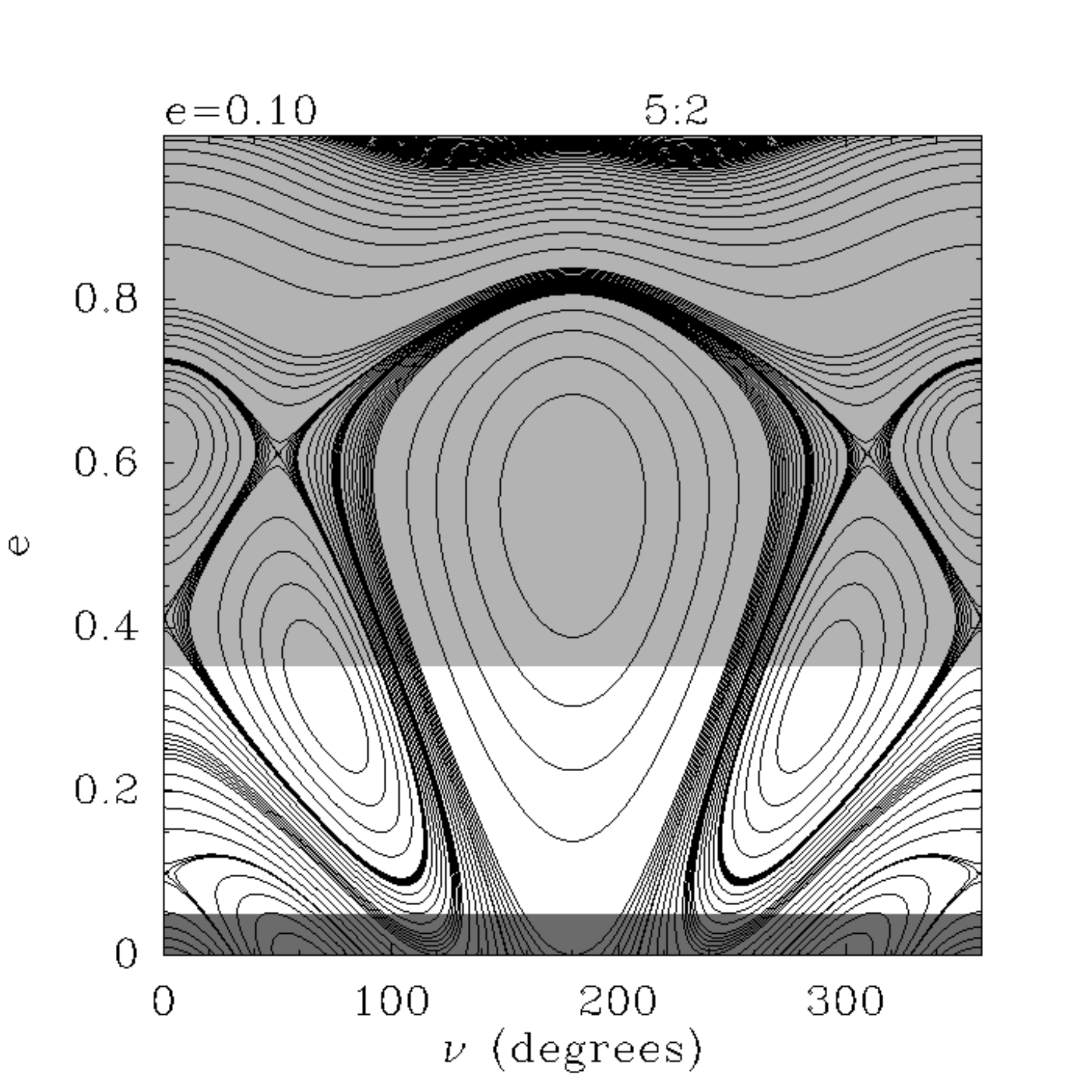}
\includegraphics[width=0.3\textwidth,height=0.3\textwidth]{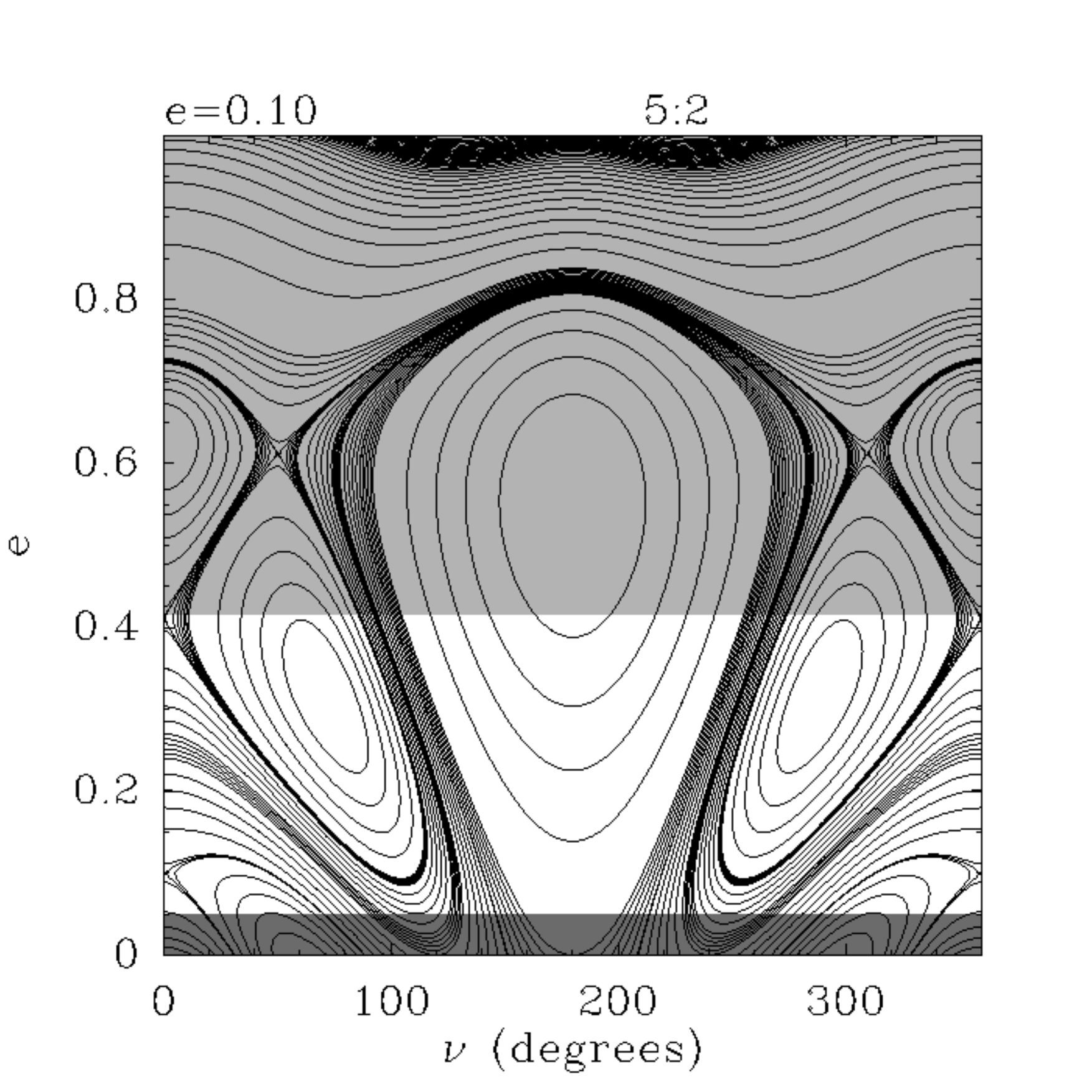}
\includegraphics[width=0.3\textwidth,height=0.3\textwidth]{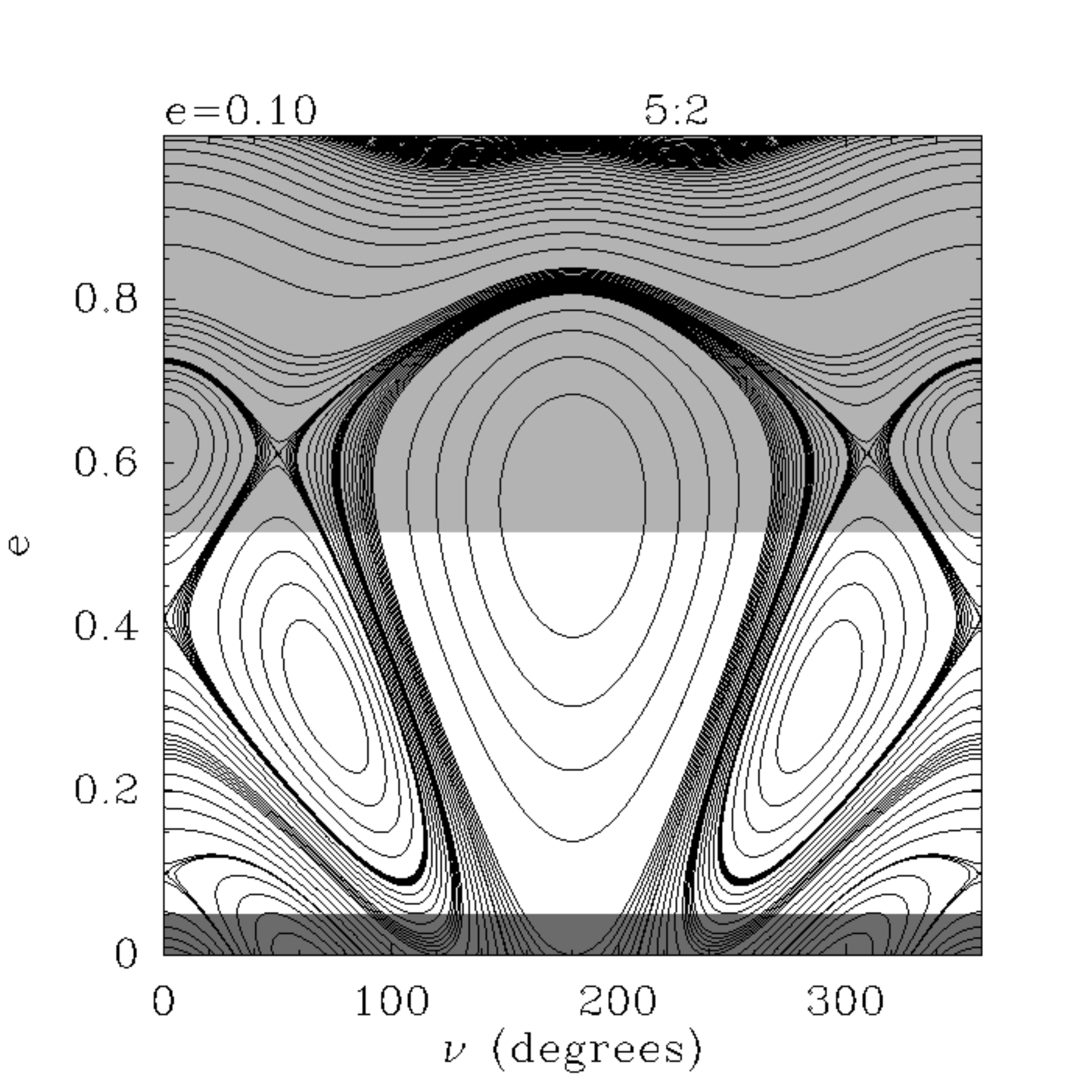}}
\makebox[\textwidth]{\includegraphics[width=0.3\textwidth,height=0.3\textwidth]{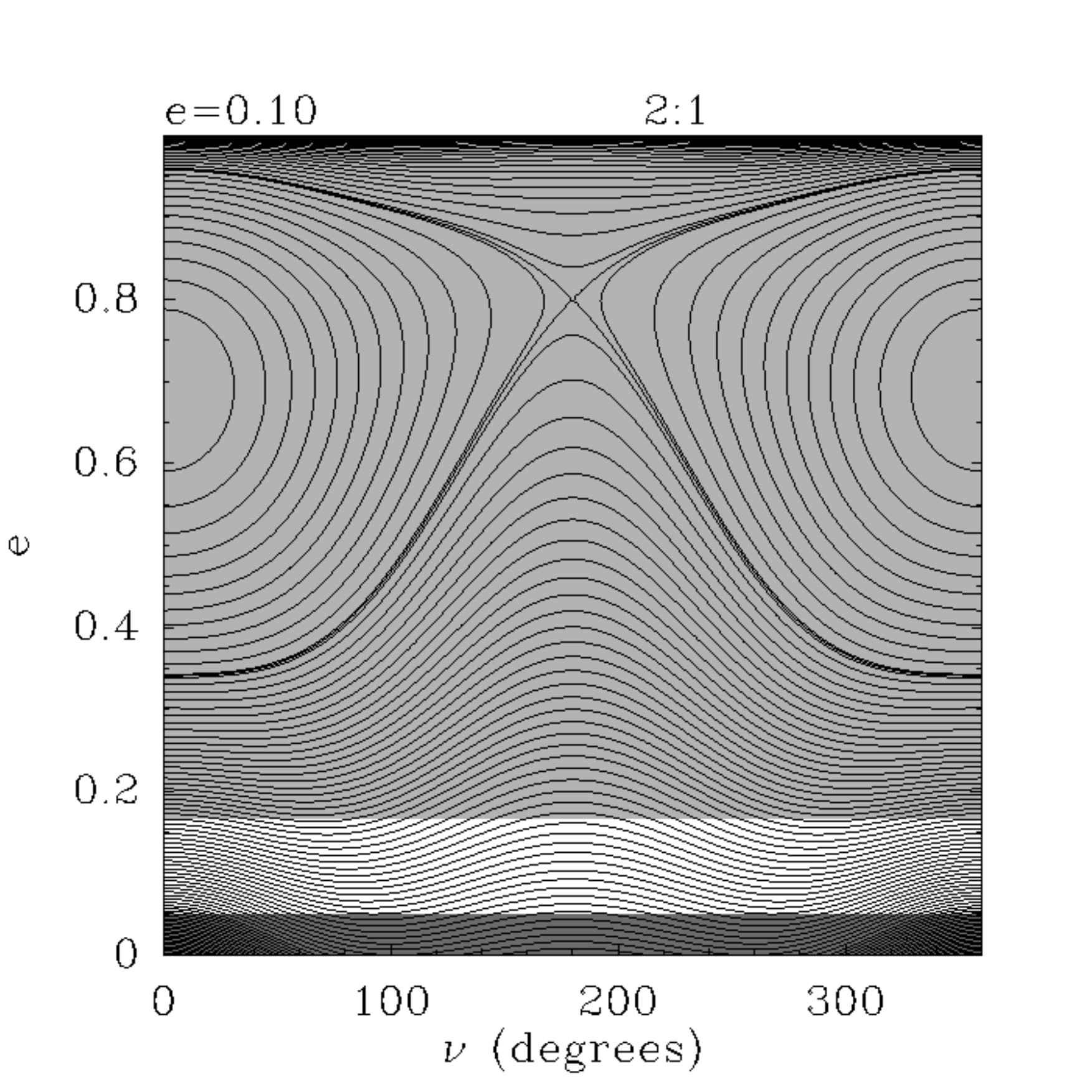}
\includegraphics[width=0.3\textwidth,height=0.3\textwidth]{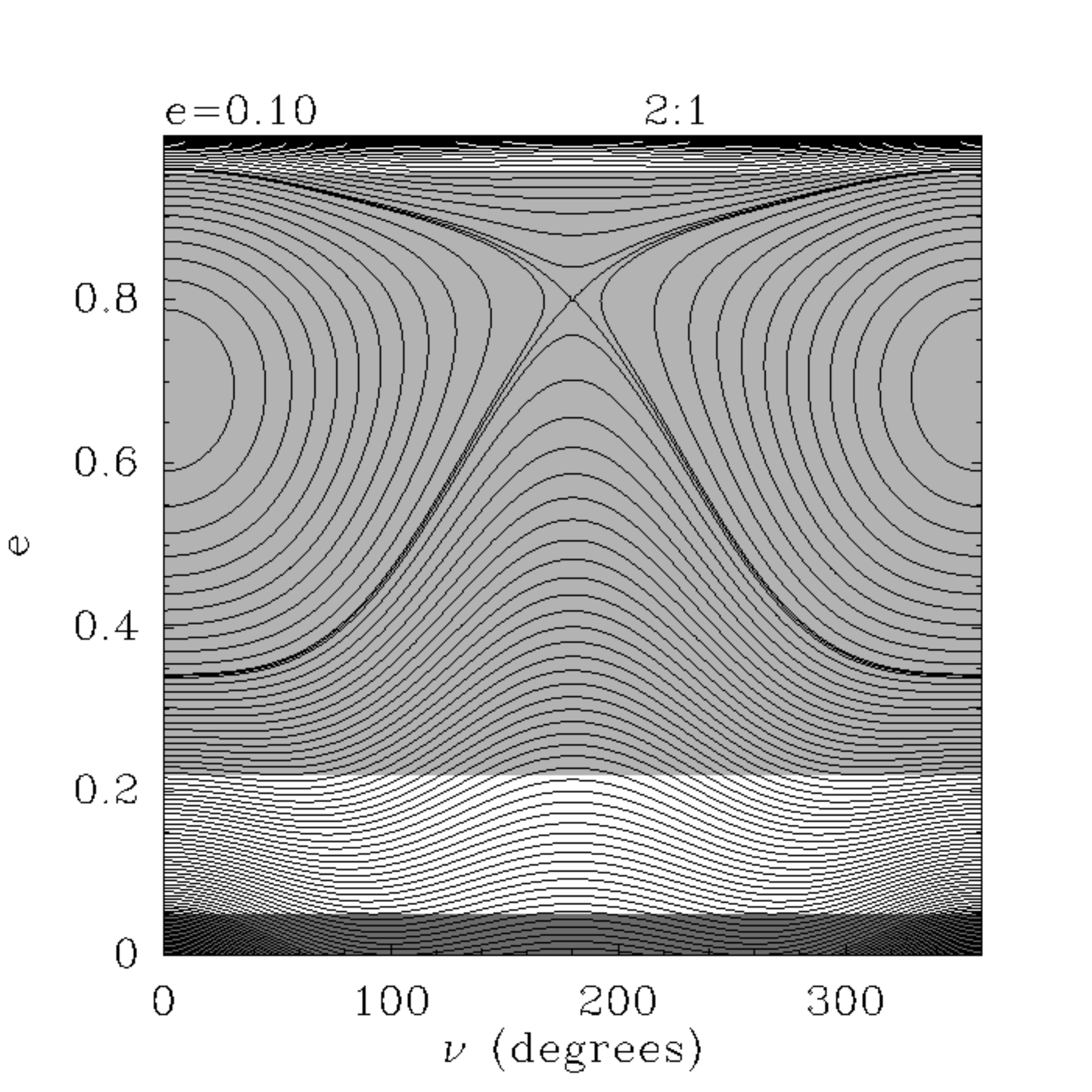}}
\caption[]{\emph{Top:} Phase diagrams for the 5:2 MMR, with Fom c of
  mass 1, 0.5, and 0.1 $\mathrm{\mjup}$, from left to
  right. \emph{Top:} Phase diagrams for the 2:1 MMR, with Fom c of
  mass 1, and 0.5 $\mathrm{\mjup}$, from left to right. Our initial
  conditions are figured in dark grey and the chaotic zone of Fom c in
  light grey. The chaotic zone of Fom c is considered to extend from
  $3.5 \mathrm{R_H}$ inner to the periastron of Fom c, to $3.5
  \mathrm{R_H}$ outer to the apastron of Fom c. Particles which start
  on a trajectory allowing it to cross the chaotic zone may then be
  scattered and set on a Fom b-like orbit.}
\label{fig:MMR_lowmass}
\end{figure*}

From Eq.~(\ref{eq:cond3}) we can see that a less massive Fom c will
require MMRs to make test-particles acquire higher eccentricities to
reach the chaotic zone of Fom c, and thus to enable the production of
Fom b-like orbits. MMRs are therefore expected to become less
efficient at producing Fom b-like orbits with decreasing mass of Fom
c.  We illustrate this aspect on Fig.~\ref{fig:MMR_lowmass}, where we
show phase-diagram of the 5:2 and 2:1 MMRs for different masses of Fom
c. In the case of the 5:2 MMR, the mass of Fom c can decrease as low
as $0.1\,\mathrm{\mjup}$ (Saturn-sized) and Fom b-like orbits are
still expected to be produced, although at a lower rate.  On the other
hand, one can see that as soon as the mass of Fom c decreases to
$1\,\mathrm{\mjup}$, the 2:1 MMR is not expected to be efficient any
longer to produce Fom b-like orbits.

Therefore, additional simulations were run for these two MMRs, identical 
to Run~F, with lower masses for Fom c (see Table~\ref{tab:initial3}).

\begin{table}[htbp]
\caption{Characteristics of initial sets of particles used to study the 
5:2 and 2:1 MMRs. Fom c itself is assumed to be orbiting Fomalhaut with 
semi-major axis $a_\mathrm{c}=107.8\,$AU and eccentricity $e_\mathrm{c}=0.1$. 
  Our sets of particles are made of 100,000 ring-like belts particles extending
  radially between boundaries given below, eccentricities randomly
  chosen between 0 and 0.05, and inclinations between 0 and $3\degr$
  relative to Fom c's orbital plane.} 
\label{tab:initial3}
\begin{tabular*}{\columnwidth}{@{\excs}lllll}
\hline\hline\noalign{\smallskip}
Run \# & Dynamical status & Semi-major axis & Theoretical resonance & 
$m_\mathrm{c}$\\
       & relative to Fom c & extent (AU) & location (AU) & $(\mathrm{\mjup})$ \\
\noalign{\smallskip}
\hline\noalign{\smallskip}
F1 & 5:2 MMR & 56.0--61.1 & 58.5 & 1.0\\
F2 & & & & 0.5 \\
F3 & & & & 0.25 \\
F4 & & & & 0.1 \\
\hline\noalign{\smallskip}
H1 & 2:1 MMR & 65.4--70.4 & 67.9 & 1.0 \\
H2 & & & & 0.5 \\
\noalign{\smallskip}\hline
\end{tabular*}
\end{table}

Interestingly, the occurrence of Fom b-like orbits is delayed by $\sim
30\,$Myr in both MMRs with a $1\,\mathrm{\mjup}$ Fom c.  This reflects
the fact that a less massive Fom c increases dynamical timescales, and
in particular, the timescale necessary for the test-particles to reach
a sufficient eccentricity to cross the chaotic zone, and be scattered.
We summarize in Table~\ref{tab:probaMMRvs_mass} the proportion of the
100,000 test-particles of our initial sample set at least once on a
Fom b-like orbit, and the probabilities that characterise the
production of Fom b-like orbits.

\begin{table}[htbp]
\caption{For each individual run that produced Fom b-like orbits,
  probability $P_\mathrm{Fomb}$ for being set on a Fom b-like orbit,
  that is, the proportion of the 100,000 test-particles of our initial
  sample set at least once on a Fom b-like orbit, average time
  $\bar{t}_\mathrm{Fom b}$ spent by these test-particles in this
  configuration, probability $P_\mathrm{>10\,Myr}$ for a Fom b-like
  orbit to have a lifetime greater than 10 Myr and probability
  $P_\mathrm{orient}$ for a Fom b-like orbit to have an orientation
  comparable to that of Fom b. We indicate as well any delay in the
  generation of Fom b-like orbits.}
\label{tab:probaMMRvs_mass}
\begin{tabular*}{\columnwidth}{@{\excs}lllllll}
\hline\hline\noalign{\smallskip}
MMR & $m_\mathrm{c}$ & $P_\mathrm{Fomb}$ & $\bar{t}_\mathrm{Fom b}$ &
$P_\mathrm{>10\,Myr}$ & $P_\mathrm{orient}$ & Delay \\
 & $(\mathrm{\mjup})$ & (\%) & (\%) & (\%) &  (Myr) \\
\noalign{\smallskip}
\hline\noalign{\smallskip}
     & 3 & $20.1$ & 1.6 & 3.5 & 15.2 & $\sim 1$ \\
 2:1 & 1 & $1.1\times 10^{-1}$ & 1.4 & 1.8 & 41.9 & $\sim 30$ \\
     & 0.5 & 0 & - & - & - & -\\
\noalign{\smallskip}\hline
     & 3   & $3.8$ & 1.2 & 2.4 & 17.6 & $\sim 2$ \\
 5:2 & 1   & $6.3\times 10^{-2}$ & 0.9 & 0 & 38.7 & $\sim$ 30--40 \\
     & 0.5 &  0 & - & - & - \\
     & 0.1 &  0 & - & - & -   \\
\noalign{\smallskip}\hline
\end{tabular*}
\end{table}

For both MMRs, the proportion of particles set on a Fom b-like orbits
is sharply decreasing with the mass of Fom c and is almost zero with
$1\,\mathrm{\mjup}$. The time spent in average by a test-particle on
its Fom b-like orbit is very short $\sim 1\,$Myr. In addition, the Fom
b-like orbits produced via the 5:2 MMR when $m_\mathrm{c}=
1\,\mathrm{\mjup}$ are highly unstable (none of them survived longer
than 10 Myr). However, since the production of Fom b-like orbits is
delayed, test-particles may be set on a Fom b-like orbit later than
the 100 Myr of the run, and test-particles on a Fom b-like orbit at
the end of the run may survive longer in this
configuration. Therefore, these quantities are to be considered with
caution and as lower limits.

No test-particle was set on a Fom b-like orbit for masses below
$1\,\mathrm{\mjup}$, which was clearly expected in the case of the 2:1
MMR, but not in the case of the 5:2, for which particles are still
expected to be able to cross the chaotic zone and Fom b-like orbits to
be produced. Again, this feature may be due to the fact that the delay
in the production of Fom b-like orbits is expected to increase when
the mass of Fom c decreases. In other words, these orbits may start
being produced after the 100 Myr of the run when $m_\mathrm{c}=
0.5\,\mathrm{\mjup}$. Therefore, we run again the simulations F1--F4
and extend the runs over 500 Myr in order to test this hypothesis.

In the case where $m_\mathrm{c}= 1\,\mathrm{\mjup}$, the proportion of
test-particles set on a Fom b-like orbit increases actually up to 1\%
over 500 Myr.  As we were expecting, Fom b-like orbits can be produced
via the 5:2 MMR when $m_\mathrm{c}= 0.25$--$0.5\,\mathrm{\mjup}$, and
their production is very interestingly delayed on timescales
comparable to the age of the system (see
Table~\ref{tab:probaMMRvs_mass}).

For $m_\mathrm{c}= 0.5-1\,\mathrm{\mjup}$, the time spent in average
by a test-particle in a Fom b-like orbit configuration has increased
up to $\sim 6\,$Myr. This time is smaller for $m_\mathrm{c}=
0.25\,\mathrm{\mjup}$, however, as mentionned, Fom b-like orbits
started to be produced very late.

\begin{table}[htbp]
\caption{Case of the 5:2 MMR with 0.1--0.5--1 $\mathrm{\mjup}$. For
  each individual run that produced Fom b-like orbits, probability
  $P_\mathrm{Fomb}$ for being set on a Fom b-like orbit, that is,
  the proportion of the 100,000 test-particles of our initial sample
  set at least once on a Fom b-like orbit, average time
  $\bar{t}_\mathrm{Fom b}$ spent by these test-particles in this
  configuration, probability $P_\mathrm{>10\,Myr}$ for a Fom b-like
  orbit to have a lifetime greater than 10 Myr and probability
  $P_\mathrm{orient}$ for a Fom b-like orbit to have an orientation
  comparable to that of Fom b. We indicate as well any delay in the
  generation of Fom b-like orbits. }
\label{tab:probaMMR52ext}
\begin{tabular*}{\columnwidth}{@{\excs}llllll}
\hline\hline\noalign{\smallskip}
 $m_\mathrm{c}$ & $P_\mathrm{Fomb}$ & $\bar{t}_\mathrm{Fom b}$ & 
$P_\mathrm{>10\,Myr}$ & $P_\mathrm{orient}$ & Delay \\
 $(\mathrm{\mjup})$ & (\%) & (Myr) & (\%) & (\%) & (Myr)  \\
\noalign{\smallskip}
\hline\noalign{\smallskip}
 1   & 1.2 & 5.7 & 10.1 & 14.2 & $\sim$ 30--40  \\
 0.5 & 0.16 & 6.4 & 7.5 & 20.2 & $\sim$ 100--150 \\
 0.25 &  $1.6\times 10^{-2}$ & 3.5 & 12.5 & 49.2 & $\sim$ 350 \\
 0.1 &  0 & - & - & - & -  \\
\noalign{\smallskip}\hline
\end{tabular*}
\end{table}

Note that while varying the mass of Fom c in our simulations, we kept
the same semi-major axis value, although the constraint for Fom c to
shape the inner edge of the outer belt at 133 AU involves that this
semi-major should increase with decreasing mass of Fom c. However, as
we have seen, the capacity of a perturber to bring test-particles in
its chaotic zone via MMR does not depend on the semi-major axis of the
perturber, and therefore, our results would still be valid if we
applied the constraint mentionned above. The only effect that a
greater semi-major axis would have is to increase the dynamical
timescales, and thus, our results are all the more valid.

The production of Fom b-like orbits via the 2:1 MMR is extremely
sensitive to the mass of Fom c and it appears not to be the most
probable origin of Fom b in our scenario. The best candidate is
therefore the 5:2 MMR, which is much less sensitive to the mass of Fom
c in its production of Fom b-like orbits, and therefore a more robust
route for Fom b to have been set on its current orbit. Moreover, this
mechanism as produced by a $0.25$--$0.5\,\mathrm{\mjup}$ Fom c can delay
the apparition of Fom b-like orbits on timescales comparable to the
age of the system, while increasing their lifetime. A lower mass limit
of $0.1\,\mathrm{\mjup}$ on the belt-shaping Fom c can be set. These
timescales are more in accordance with our witnessing of the orbit of
Fom b. Moreover, a $0.25$--$0.5\,\mathrm{\mjup}$ Fom c would allow Fom b
not to be ejected too quickly from its present-day orbit, as
underlined by \citet{2014A&A...561A..43B}. Finally, a
$0.25-0.5\,\mathrm{\mjup}$ Fom c is completely in accordance with the
shaping the outer belt into the observed eccentric ring, as shown by
\citet{2006MNRAS.372L..14Q}.

\subsection{Preferential apsidal orientation}\label{sec:apsidal}

A notable feature of our results is that the Fom b-like orbits formed
tend to be apsidally aligned with the orbit of Fom c in a very general
manner, even when these originated directly from the chaotic zone of
Fom c, where they were expected to suffer random encounters and thus
be put on randomly apsidally aligned Fom b-like orbits. This hints at
the fact that the whole dynamical process of production of Fom b-like
orbits is more complex than previously thought.  We have so far
proposed a two-steps scenario, where a test particle firstly reaches
the chaotic zone of Fom c on timescales comparable to the age of the
system via a MMR mechanism with Fom c, and where this test-particle
secondly suffers a close-encounters with Fom c.

However, a closer study of the whole dynamical behaviour of a
test-particle along the two-steps process that we have proposed,
and in particular an exam of the orbits resulting from
 close-encounters with Fom c, shows that an additional third step
involving secular interactions with Fom c is not only required, but
also explain the tendency for apsidal alignement.
\subsubsection{Close-encounters with Fom c}
Close-encounters can be investigated analytically in a very
simple manner considering the Tisserand parameter $C_\mathrm{T}$ of
a test particle. If we assume here coplanarity between Fom c and the
test-particle, this quantity reads
\begin{equation}
C_\mathrm{T} = \frac{a_\mathrm{c}}{a} 
+ 2\sqrt{\frac{a}{a_\mathrm{c}}}\sqrt{1-e^2} \qquad,
\label{tisserand}
\end{equation} 
where $a_\mathrm{c}$ is the semi-major axis of Fom c, and where $a$ and $e$ are the semi-major axis and eccentricity of the test-particle.

Tisserand parameter is closely related to the Jacobi invariant which
is a conserved quantity in the framework of the circular restricted
3-body system, even after close encounters. Here the perturber (Fom c)
has moderate but non-zero eccentricity. Strictly speaking,
$C_\mathrm{T}$ is thus not conserved, but detailed studies focusing on
Jupiter perturbed comets showed that in most cases, $C_\mathrm{T}$
remained preserved within $\sim 1\,\%$ despite the eccentricity of
Jupiter \citep{1995EM&P...68...71C}. Here the assumed eccentricity (0.1) is
only twice that of Jupiter, so that we expect $C_\mathrm{T}$ to be
perserved within a few percents in close encounters. This accuracy is
sufficient for our analysis.

Consider a particle initially locked in a MMR with Fom c, having a
neary constant semi-major axis $\mathrm{a_{MMR}}$ and a growing
eccentricity. Assume it has reached eccentricity $e$ when crossing the
chaotic zone. Then it suffers one or many close encounter episodes
with Fom c. Afterwards, its semi-major axis $a$ and eccentricity $e'$
are related to $\mathrm{a_{MMR}}$ and $e$ by the conservation of the
Tisserand parameter:
\begin{equation}
\frac{a_\mathrm{c}}{a_\mathrm{MMR}} + 2\sqrt{\frac{a_\mathrm{MMR}}{a_\mathrm{c}}}\sqrt{1-e^2} = \frac{a_\mathrm{c}}{a^\prime} + 2\sqrt{\frac{a^\prime}{a_\mathrm{c}}}\sqrt{1-e^{\prime 2}} \qquad.
\label{tisserand1}
\end{equation} 
Depending on the resonance considered, there are constraints on
$a_\mathrm{MMR}$ and $e$ for the orbit to be able to cross the
chaotic zone. For instance, in the case of the 5:2 MMR with a
$3\,\mathrm{\mjup}$ Fom c, we must have $0.2\la e\la 0.8$. This
naturally translates to constraints on $a'$ and $e'$ via
Eq.(\ref{tisserand1}).  Note that these constraints depend on the mass
of Fom c, since this parameter controls the width of the chaotic zone
and thus the values of eccentricities allowed to the test-particles.

\begin{figure*}[htbp]
\makebox[\textwidth]{\includegraphics[width=0.5\textwidth]{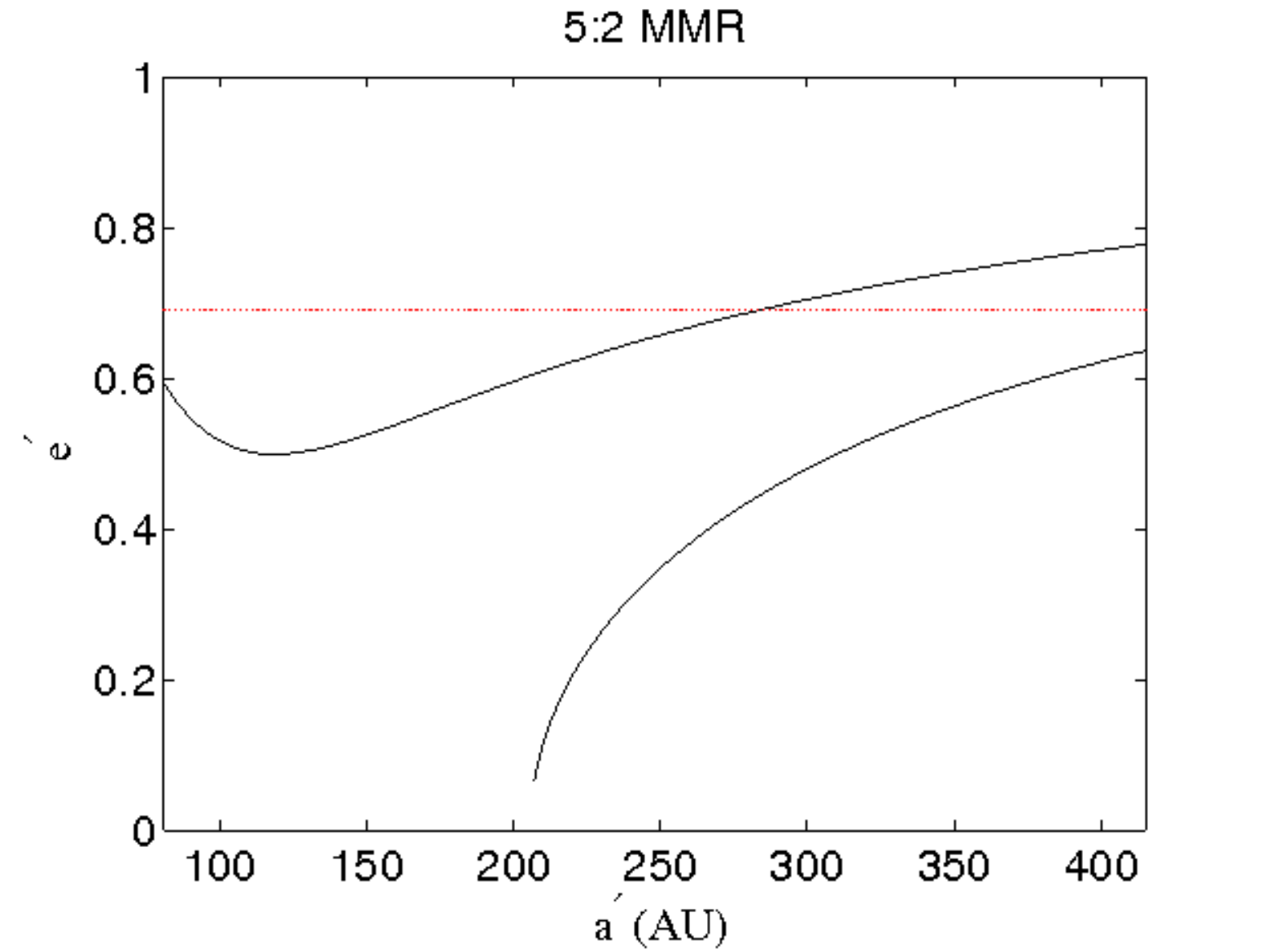}
\includegraphics[width=0.5\textwidth]{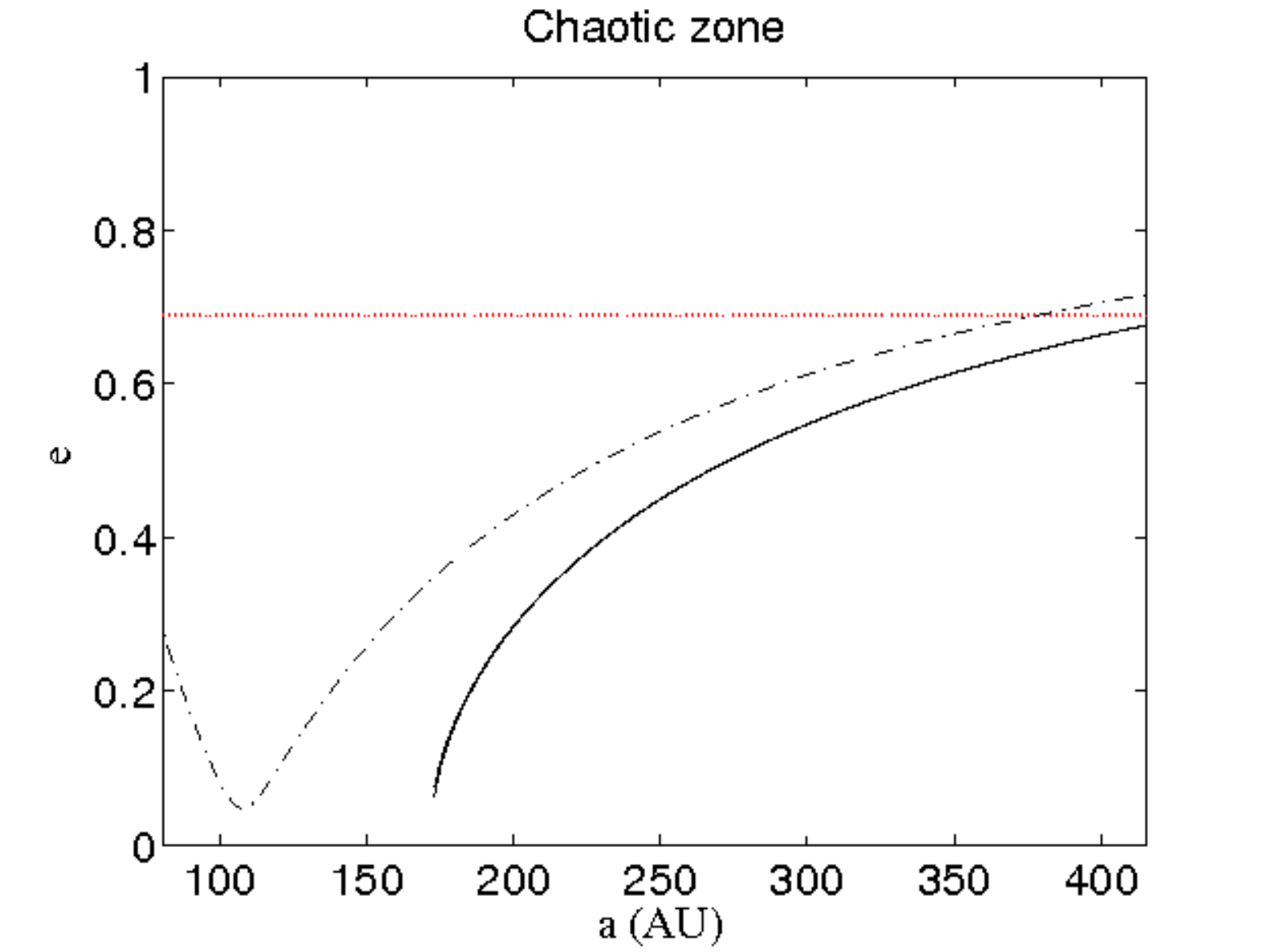}}
\caption[]{Theoretical eccentricities $e^\prime$ adopted by
  test-particles after a close-encounter with Fom c which has set them
  on an orbit with semi-major axis compatible with the orbit of Fom
  b. On the left panel, the 5:2 MMR will constrain the semi-major axis to $a_{\mathrm{MMR}}$ and will allow a test-particle to cross the chaotic zone of Fom c for values of eccentricity $e$ between $0.2$ and $0.8$, hence the configurations $(a',e')$ allowed to a particle after its encounter with Fom c are comprised between two curves. On the right panel, close-encounters occur in the chaotic zone of Fom c, with initially low-eccentricity particles $(0<e<0.05)$, but here the semi-major axis can span values from the inner edge of the chaotic zone to the semi-major axis of Fom c. Therefore, there is a total of four curves on this plot, two curves for each boundary value in semi-major axis, but due to the small span in eccentricity, these are very close and appear as a single one. The horizontal red dotted line figures the minimum eccentricity
  required for an orbit to be compatible with this of Fom b.}
\label{fig:postenc}
\end{figure*}

Constraints on $a^\prime$ can also be derived via our definition of 
a Fom b-like orbit, namely 
$81,\mathrm{AU}\le\mathrm{a^\prime}\le 415\,\mathrm{AU}$.
Once this constraints are incorporated into Eq.~\ref{tisserand1}, this leads 
to constraints on the eccentricity $e^\prime$ that the test-particle
 can have after the close-encounter and when having a semi-major axis 
compatible with the definition of a Fom b-like orbit:
\begin{equation}
e^\prime = \left[ 1 - \frac{1}{4} \left( C_\mathrm{T}-\frac{a_\mathrm{c}}{a^\prime} \right)^2 \frac{a_\mathrm{c}}{a^\prime} \right]^{1/2} \qquad.
\end{equation}
This resulting possible eccentricities after a close-encounter are
displayed in Fig.~\ref{fig:postenc} for the 5:2 MMR and for the
chaotic zone. In the chaotic zone case, the limits on $e$ are
simply the limits set by our initial conditions on the eccentricity of
the test-particles, that is, $e\le 0.05$.

Figure~\ref{fig:postenc} reveals that the eccentricity after the
scattering event(s) rarely exceeds $\sim 0.6$--$0.7$, whereas the minimum
eccentricity required for the orbit to be fully qualified of Fom
b-like is $0.69$. It thus seems that directly generating Fom b like
orbits from (even multiple) close encounters is difficult. But, as we
detail it below, secular evolution after the close encounter episode
can help moving to higher eccentricies and also provide explanation
for the apsidal alignment with Fom c.
\subsubsection{Further secular evolution with Fom c}
Particles initially locked in a MMR with Fom c, and that have
undergone a close encounter episode keep being perturbed in a secular
manner with Fom c even after the last encounter. This behaviour can be
investigated semi-analytically in a similar way as we did in the
resonant case in Sect.~2 (Fig.~2). Now, as the particle is no longer
locked in a MMR with Fom c, it secular motion can be described
performing a double average of the interaction Hamiltonian over both
orbits (see background theory in Beust et al. 2014).
\begin{figure}[htbp]
\resizebox{\hsize}{!}{
\includegraphics[width=0.45\textwidth,height=0.35\textwidth]{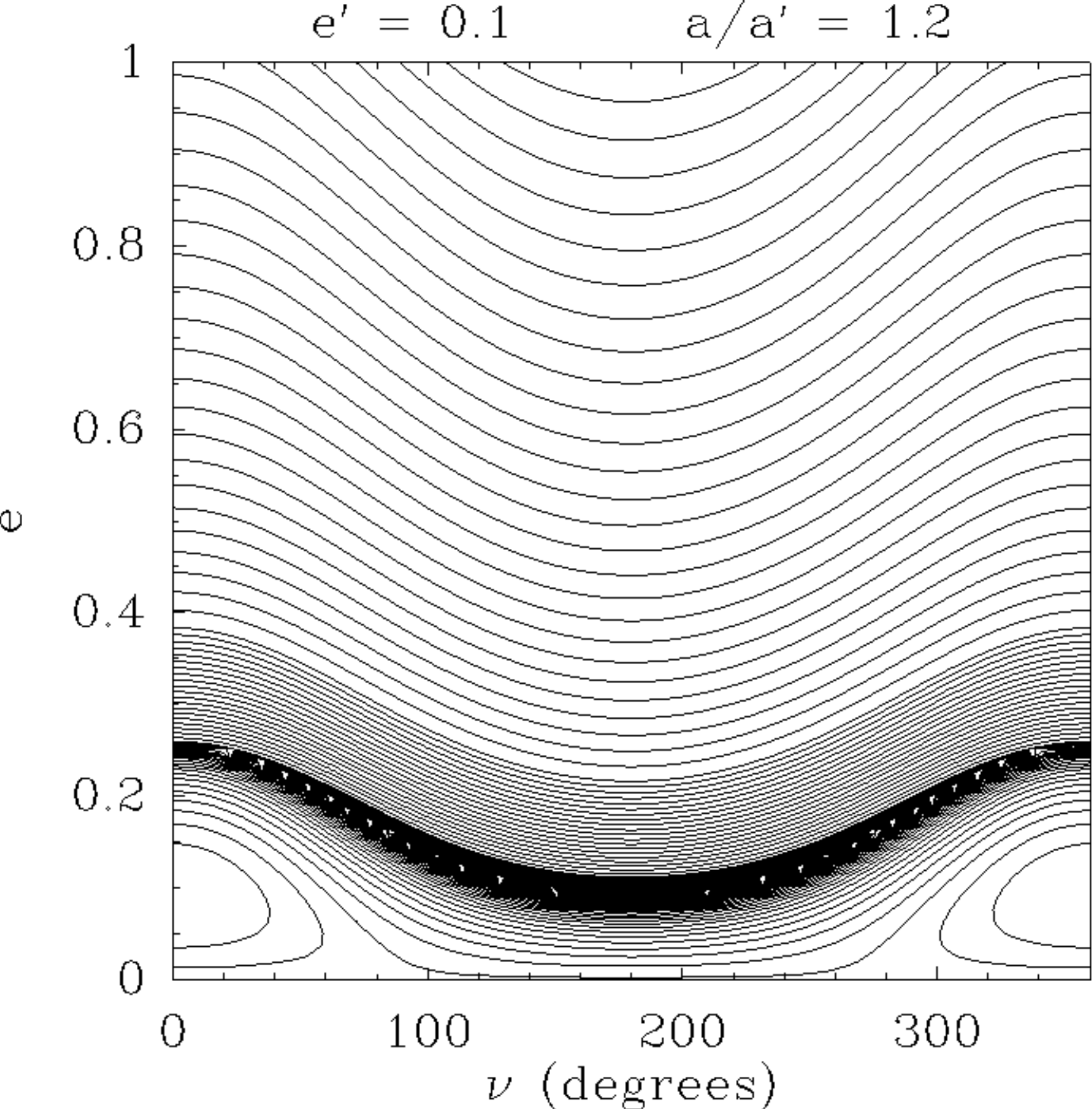}}
\caption[]{Example of secular evolution of a test-particle under the 
dynamical influence of Fom c, for a typical semi-major axis ratio 
of $a/a_\mathrm{c}=1.2$.}
\label{fig:sec_evol}
\end{figure}
This is illustrated in Fig.~\ref{fig:sec_evol}, which shows a phase
diagram of this secular Hamiltonian for a particle having
$a'/a_\mathrm{c}=1.2$, assuming coplanarity of both orbits and
$e_\mathrm{c}=0.1$. Following Fig.~\ref{fig:postenc}, let us assume that after
the close encounter episode, the particle appears in this diagram at
$e'\simeq 0.7$. Then is further secular evolution can be readily seen
of Fig.~\ref{fig:sec_evol} following the Hamiltonian level curve it
appears on. It actually depends on the starting value of $\nu$. If the
particle stars at $\nu\simeq 0$, the secular evolution will cause its
eccentricity to first decrase and in any case never overcome the
starting eccentricity. This particle will never reach a Fom b-like
orbit. Conversely, a particle starting at $\nu\simeq 180\degr$ will
undergo a secular eccentricity increase that will drive it above
$e'=0.8$ near $\nu=0$. At this point the particle has now reached a
Fom b-like configuration. But $\nu\simeq 0$ exactly means apsidal
alignment. The key point here is that in the level curves of
Fig.~\ref{fig:sec_evol}, the maximum eccentricity is reached at
$\nu=0$. This description is actually a high eccentricity equivalent
to the analytical pericenter glow theory described by Wyatt
(2005). According to this scenario, low eccentricity particles
perturbed by a low eccentricity planet undergo a secular eccentricity
evolution where the maximum eccentricity is reached together with
apsidal alignment (see Wyatt 2005 and Beust 2014 for details). This
configuration corresponds indeed to the bottom curves of
Fig.~\ref{fig:sec_evol}. Here our Fom b progenitors move at large
eccentricity on the upper curves of Fig.~\ref{fig:sec_evol}, so that a
full analytical formulation of the motion is not possible. But the
qualitative result remains: the maximum eccentricity is reached for
$\nu=0$.

So, our three steps scenario is now the following: Particles trapped
in MMRs with Fom c first undergo a resonant eccentricity increase at
$\sim$constant semi-major axis up to a point they cross the chaotic
zone. Then in a second phase they have one or several close encounters
with Fom c that extract them from the MMR and drastically change their
semi-major axes, bringing them to a $a'$ value compatible with Fom
b-like orbits and to $e'\simeq0.7$. In the third phase, they keep
being secularly perturbed by Fom c at constant $a'$, while their
eccentricities fluctuate. The particles starting the third phase close
to $\nu\simeq 0$ keep evolving below $e'\simeq0.7$ and never reach a
Fom b-like state. But hose appearing at $\nu\simeq 180\degr$ undergo a
further eccentricity evolution above $e'\simeq0.7$ that drives them to
Fom b-like orbits when $\nu=0$ is reached. We claim that Fom b could
be one of these particles, initially originating from an inner MMR
(typically the 5:2 one which is among the most efficient ones), and
now having reached $e'\ga 0.8$ and apsidal alignment (i.e., $\nu\simeq
0$) with Fom c.

\begin{figure*}[htbp]
\makebox[\textwidth]{\includegraphics[width=0.75\textwidth]{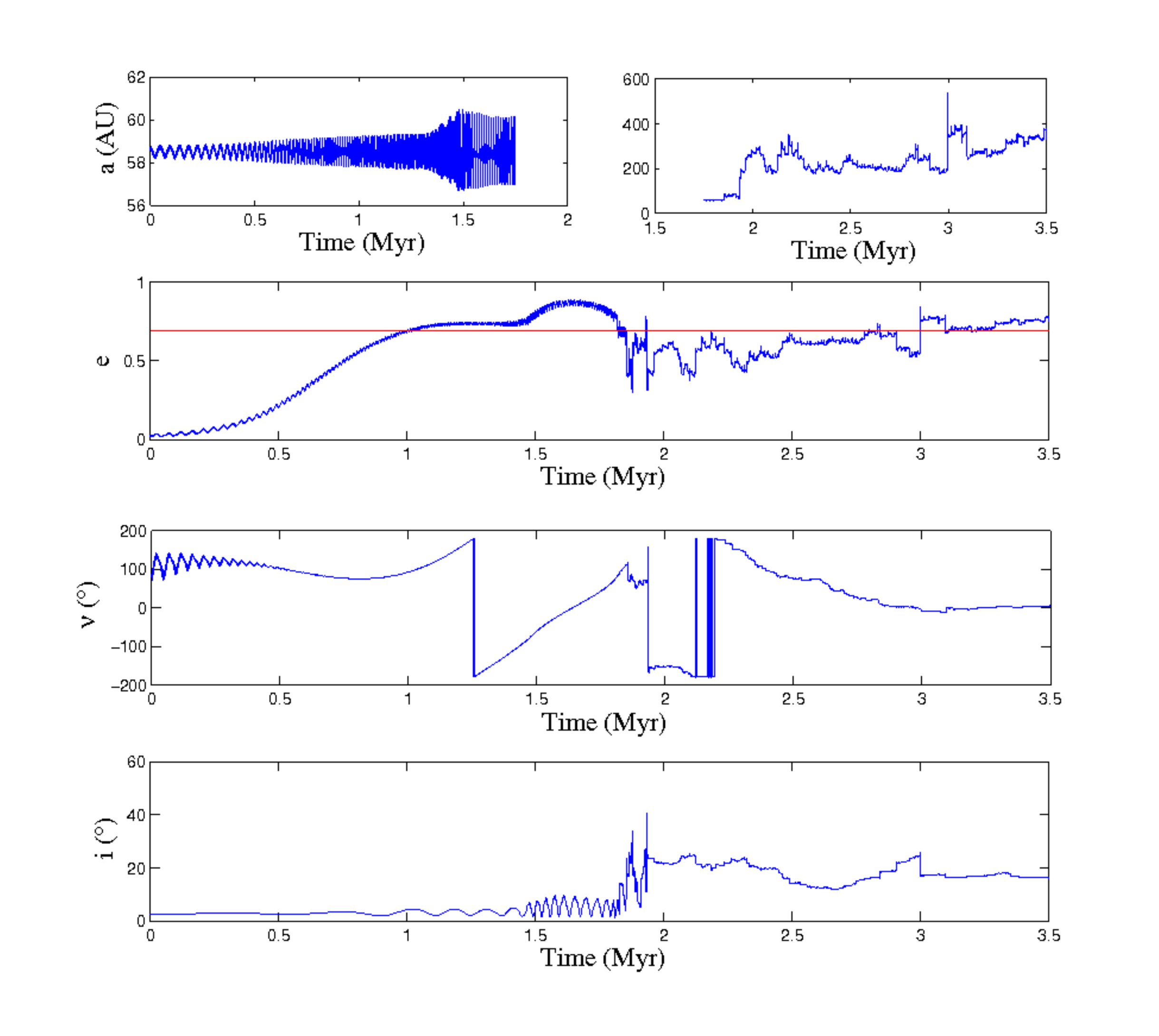}}
\caption[]{Example evolution of the semi-major axis, eccentricity,
  longitude of periastron and inclination of a test-particle set on a Fom b-like orbit
  via the 5:2 MMR route, from top to bottom, respectively. The
  semi-major axis evolution is splitted into the resonant regime, on
  left, and the secular regime, on right.  During its resonant
  evolution, the test-particle endures only small variations of its
  semi-major axis, while its eccentricity increases. Its suffers a
  close encounter with Fom c at high eccentricity because its orbits
  crosses the chaotic zone of Fom c. After the close encounters, its
  semi-major axis is compatible with that of Fom b. However, its
  eccentricity is not, as is figured by the horizontal red line which
  indicates the minimum eccentricity required for an orbit to be
  compatible with that of Fom b. The eccentricity gradually increases
  due to secular evolution and finally reaches Fom b compatible values
  at $\sim 3\,$ Myr with a $3\,\mjup$ Fom c. With a Neptune-Saturn mass Fom c, this timescale increases up to values of the order of several 100 Myr. As can be seen on the bottom-panel, the
  evolution of the eccentricity is accompanied by an evolution of the
  longitude of periastron, which tends to zero, and thus, to an
  apsidal alignement of the orbit with this of Fom c.}
\label{fig:process}
\end{figure*}

Figure~\ref{fig:process} exactly illustrates this three steps
scenario. It shows the semi-major axis, eccentricity and longitude or
periastron secular evolution of one particle extracted from our
simulation, initially trapped in 5:2 MMR with Fom c. Up to $\sim
1.8\,$Myr the particle remains in the resonce while its eccentricity
increases. Then it enters a chaotic phase characterized by encounters
with Fom c. After $\sim 2.2\,$Myrs, there are no more encounters, but
the particle keeps being secularly perturbed by Fom c. Starting this
third phase at $e'\simeq 0.6$ and $\nu\simeq 180\degr$, it evolves
towards larger eccentricities and $\nu=0$. After $\sim 3\,$Myrs it has
reached a Fom b-like state.


\section{Conclusion}\label{sec:conclusion}
The scattering events generating orbits fully comparable to the orbit of Fom b, either in terms of semi-major axis and eccentricity, but also in near-coplanarity and apsidal alignement with the belt-shaping putative Fom c see Table~\ref{tab:constraints} and Fig.~\ref{fig:fombALL_orient}), is a very robust mechanism when generated by a perturber with an eccentricity 0.1, whether these orbits are primarily due to a scattering events, or secondarily, via a MMR. 
However, MMRs are the most probable route for Fom b to have been set on its current orbit in our scenario. Indeed, primary scattering events scatter the material out of the system on timescales much shorter than the age of the system, while MMRs tend to delay the production of Fom b-like orbits, potentially on timescales comparable to the age of the Fomalhaut system. 
This delay increases with decreasing mass of Fom c, and so does the average lifetime of Fom b-like orbits. However, the ability of a MMR to bring test-particles in the chaotic zone of Fom c, and thus the efficiency of a MMR to produce Fom b-like orbits, decreases with decreasing mass of Fom c. Therefore, the mass of Fom c should be sufficient for a given MMR to produce Fom b-like orbits, but should not exceed a given value in order for the production of Fom b-like orbits to be delayed by timescales compatible with its detection at the age of the system. The MMR which realises this compromise the best is the 5:2 MMR. New constraints on the mass of the unseen Fom c in our scenario are $\mathrm{m_c}= 0.25-0.5\,\mathrm{\mjup}$. These constraints are compatible with the witnessing of a transient planetary system configuration where the orbits of Fom b and Fom c cross each other that is sufficiently stable to be witnessed \citep{2014A&A...561A..43B}, and observational constraints.
Regarding the distance of the putative Fom c from its host star, the prospects for detecting it rely on direct imaging. However, if new generation instruments such as VLT-SPHERE and Gemini-GPI are expected to offer direct detection up to 100 AU from the host star, these detections will be limited to planets with masses comparable to or larger than this of Jupiter. Therefore, prospects for detecting the putative Fom c remain extremely limited with currently available instruments.
Finally, it is also crucial that MMRs are generated by a perturber with an eccentricity $\sim 0.1$ such as this of Fom c in order to produce Fom b-like orbits.
These constraints are fully compatible with the shaping of the outer belt \citep{2006MNRAS.372L..14Q}.
Considering that it would have been difficult to form Fom b from resonant material, since eccentricities and thus relative velocities of solids are increased, which thus challenges their accretion, it is most probable that there were migration processes in this system. Fom b and/or Fom c are very likely to have migrated in order for Fom b to find itself at a MMR location. An outward migration process has been put forward to explain the presence of the belt-shaping planet at a distance of the order of 100 AU from its host star by \citet{2009ApJ...705L.148C}. This mechanism implies migration of a pair of planets in MMR: if the inner planet is more massive than the outer one, both planets can migrate outwards in a common gap in the original gaseous protoplanetary disk. However, the eccentricity of these planets are excited by their MMR configuration, but also damped by the gaseous disk \citep{2008A&A...483..325C}. In \citet{2009ApJ...705L.148C}, this resulted into planets with orbital eccentricities too moderate ($\sim 0.02-0.03$) to be compatible with that of the belt-shaping planet. Planetesimal-driven migration at later stages of the system evolutions, when gas has dissipated, could both explain the outward migration of the belt-shaping planet and its orbital eccentricity, since the abscence of gas prevents orbital eccentricities to be damped during this migration process. However, as for the early migration scenario, this would involve the presence of another massive body inner to the belt shaping planet, which questions the compatibility of our scenario with an additional putative Fom d.

Finally, a significant and broad population of small bodies were set on highly eccentric orbits via MMRs in our scenario. As the eccentricity of a resonant test-particles increases while its semi-major axis suffers only small relative variations, its periastron will obviously decrease. This means that if a population of small bodies was residing in the vicinity of Fom c, Fom b, or simply in MMR, a significant amount of this material has spent some time in the inner parts of the system, and this might be linked with the presence of two inner dust belts in the Fomalhaut system, a hot, very close, at $\sim 0.1-0.3\,$AU, and another, warm at about 2 AU \citep{2013A&A...555A.146L}. This will be the subject of a forthcoming paper (Faramaz et al. in prep).

Far from being paradoxal, the configuration of the Fomalhaut system is in fact logical, that is, if there are clues for a perturber on a 0.1 eccentric orbit in a system, bodies on Fom b-like orbits should be expected to be present in the system, in a continuous way as long as material is available either in the chaotic zone or at MMR locations, and also once a given MMR starts producing Fom b-like orbits, which can be delayed very late in the life of a system.
This suggests that warm and hot inner dusty belts potentially resulting from this process may start to be produced very late in the history of a system. In the same manner that it might explain the presence of inner dust belts in the Fomalhaut system, this may also give a solution to the yet unexplained detection of numerous hot belts in systems older than 100 Myr, and which contain levels of dust surprisingly large at such ages \citep[12 to 30\% of stars]{2013A&A...555A.104A,2014arXiv1409.6143E}.
\citet{2012A&A...548A.104B} and \citet{2014arXiv1404.2606B} have respectively investigated whether scattering of planetesimals by a chain of planets or subsequent to planetary migration, as possible mechanisms to explain the presence of such hot belts over several 100 Myr. From the three-step process revealed in this paper, one should not necessarily assume that hot belts in systems older than 100 Myr have been sustained over the system's age, and suggests that some of these hot belts may be related to the presence of a massive and slightly eccentric planet in the system.


\section*{Aknowledgements: }

We thank the anonymous referee for helping us to clarify this paper.
The authors acknowledge the support of the ANR-2010 BLAN-
0505-01 (EXOZODI).
P.K. and J.R.G. thank support from NASA NNX11AD21G, GO-11818, GO-12576, and NSF AST-0909188.
Computations presented in this paper were performed at the Service 
Commun de Calcul Intensif de l'Observatoire de Grenoble (SCCI) on the 
super-computer funded by the Agence Nationale pour la Recherche under 
contracts ANR-07-BLAN-0221, ANR-2010-JCJC-0504-01 and ANR-2010-JCJC-0501-01.

\bibliography{biblio}

\begin{thebibliography}{41}
\expandafter\ifx\csname natexlab\endcsname\relax\def\natexlab#1{#1}\fi

\bibitem[{{Absil} {et~al.}(2013){Absil}, {Defr{\`e}re}, {Coud{\'e} du Foresto},
  {Di Folco}, {M{\'e}rand}, {Augereau}, {Ertel}, {Hanot}, {Kervella},
  {Mollier}, {Scott}, {Che}, {Monnier}, {Thureau}, {Tuthill}, {ten Brummelaar},
  {McAlister}, {Sturmann}, {Sturmann}, \& {Turner}}]{2013A&A...555A.104A}
{Absil}, O., {Defr{\`e}re}, D., {Coud{\'e} du Foresto}, V., {et~al.} 2013,
  \aap, 555, A104

\bibitem[{{Beust} {et~al.}(2014){Beust}, {Augereau}, {Bonsor}, {Graham},
  {Kalas}, {Lebreton}, {Lagrange}, {Ertel}, {Faramaz}, \&
  {Th{\'e}bault}}]{2014A&A...561A..43B}
{Beust}, H., {Augereau}, J.-C., {Bonsor}, A., {et~al.} 2014, \aap, 561, A43

\bibitem[{{Beust} \& {Morbidelli}(1996)}]{1996Icar..120..358B}
{Beust}, H. \& {Morbidelli}, A. 1996, \icarus, 120, 358

\bibitem[{{Beust} \& {Morbidelli}(2000)}]{2000Icar..143..170B}
{Beust}, H. \& {Morbidelli}, A. 2000, \icarus, 143, 170

\bibitem[{{Bonsor} {et~al.}(2012){Bonsor}, {Augereau}, \&
  {Th{\'e}bault}}]{2012A&A...548A.104B}
{Bonsor}, A., {Augereau}, J.-C., \& {Th{\'e}bault}, P. 2012, \aap, 548, A104

\bibitem[{{Bonsor} {et~al.}(2014){Bonsor}, {Raymond}, {Augereau}, \&
  {Ormel}}]{2014arXiv1404.2606B}
{Bonsor}, A., {Raymond}, S.~N., {Augereau}, J.-C., \& {Ormel}, C.~W. 2014,
  ArXiv e-prints

\bibitem[{{Carusi} {et~al.}(1995){Carusi}, {Kres{\'a}k}, \&
  {Valsecchi}}]{1995EM&P...68...71C}
{Carusi}, A., {Kres{\'a}k}, {\v L}., \& {Valsecchi}, G.~B. 1995, Earth Moon and
  Planets, 68, 71

\bibitem[{{Chiang} {et~al.}(2009){Chiang}, {Kite}, {Kalas}, {Graham}, \&
  {Clampin}}]{2009ApJ...693..734C}
{Chiang}, E., {Kite}, E., {Kalas}, P., {Graham}, J.~R., \& {Clampin}, M. 2009,
  \apj, 693, 734

\bibitem[{{Crida} {et~al.}(2009){Crida}, {Masset}, \&
  {Morbidelli}}]{2009ApJ...705L.148C}
{Crida}, A., {Masset}, F., \& {Morbidelli}, A. 2009, \apjl, 705, L148

\bibitem[{{Crida} {et~al.}(2008){Crida}, {S{\'a}ndor}, \&
  {Kley}}]{2008A&A...483..325C}
{Crida}, A., {S{\'a}ndor}, Z., \& {Kley}, W. 2008, \aap, 483, 325

\bibitem[{{Currie} {et~al.}(2013){Currie}, {Cloutier}, {Debes}, {Kenyon}, \&
  {Kaisler}}]{2013ApJ...777L...6C}
{Currie}, T., {Cloutier}, R., {Debes}, J.~H., {Kenyon}, S.~J., \& {Kaisler}, D.
  2013, \apjl, 777, L6

\bibitem[{{Deller} \& {Maddison}(2005)}]{2005ApJ...625..398D}
{Deller}, A.~T. \& {Maddison}, S.~T. 2005, \apj, 625, 398

\bibitem[{{Duncan} {et~al.}(1989){Duncan}, {Quinn}, \&
  {Tremaine}}]{1989Icar...82..402D}
{Duncan}, M., {Quinn}, T., \& {Tremaine}, S. 1989, \icarus, 82, 402

\bibitem[{{Ertel} {et~al.}(2014){Ertel}, {Absil}, {Defrere}, {Le Bouquin},
  {Augereau}, {Marion}, {Blind}, {Bonsor}, {Bryden}, {Lebreton}, \&
  {Milli}}]{2014arXiv1409.6143E}
{Ertel}, S., {Absil}, O., {Defrere}, D., {et~al.} 2014, ArXiv e-prints

\bibitem[{{Faramaz} {et~al.}(2014){Faramaz}, {Beust}, {Th{\'e}bault},
  {Augereau}, {Bonsor}, {del Burgo}, {Ertel}, {Marshall}, {Milli},
  {Montesinos}, {Mora}, {Bryden}, {Danchi}, {Eiroa}, {White}, \&
  {Wolf}}]{2014A&A...563A..72F}
{Faramaz}, V., {Beust}, H., {Th{\'e}bault}, P., {et~al.} 2014, \aap, 563, A72

\bibitem[{{Galicher} {et~al.}(2013){Galicher}, {Marois}, {Zuckerman}, \&
  {Macintosh}}]{2013ApJ...769...42G}
{Galicher}, R., {Marois}, C., {Zuckerman}, B., \& {Macintosh}, B. 2013, \apj,
  769, 42

\bibitem[{{Graham} {et~al.}(2013){Graham}, {Fitzgerald}, {Kalas}, \&
  {Clampin}}]{2013AAS...22132403G}
{Graham}, J.~R., {Fitzgerald}, M.~P., {Kalas}, P., \& {Clampin}, M. 2013, in
  American Astronomical Society Meeting Abstracts, Vol. 221, American
  Astronomical Society Meeting Abstracts, 324.03

\bibitem[{{Ida} {et~al.}(2000){Ida}, {Bryden}, {Lin}, \&
  {Tanaka}}]{2000ApJ...534..428I}
{Ida}, S., {Bryden}, G., {Lin}, D.~N.~C., \& {Tanaka}, H. 2000, \apj, 534, 428

\bibitem[{{Janson} {et~al.}(2012){Janson}, {Carson}, {Lafreni{\`e}re},
  {Spiegel}, {Bent}, \& {Wong}}]{2012ApJ...747..116J}
{Janson}, M., {Carson}, J.~C., {Lafreni{\`e}re}, D., {et~al.} 2012, \apj, 747,
  116

\bibitem[{{Kalas} {et~al.}(2008){Kalas}, {Graham}, {Chiang}, {Fitzgerald},
  {Clampin}, {Kite}, {Stapelfeldt}, {Marois}, \& {Krist}}]{2008Sci...322.1345K}
{Kalas}, P., {Graham}, J.~R., {Chiang}, E., {et~al.} 2008, Science, 322, 1345

\bibitem[{{Kalas} {et~al.}(2005){Kalas}, {Graham}, \&
  {Clampin}}]{2005Natur.435.1067K}
{Kalas}, P., {Graham}, J.~R., \& {Clampin}, M. 2005, \nat, 435, 1067

\bibitem[{{Kennedy} \& {Wyatt}(2011)}]{2011MNRAS.412.2137K}
{Kennedy}, G.~M. \& {Wyatt}, M.~C. 2011, \mnras, 412, 2137

\bibitem[{{Kenyon} {et~al.}(2014){Kenyon}, {Currie}, \&
  {Bromley}}]{2014ApJ...786...70K}
{Kenyon}, S.~J., {Currie}, T., \& {Bromley}, B.~C. 2014, \apj, 786, 70

\bibitem[{{Kirsh} {et~al.}(2009){Kirsh}, {Duncan}, {Brasser}, \&
  {Levison}}]{2009Icar..199..197K}
{Kirsh}, D.~R., {Duncan}, M., {Brasser}, R., \& {Levison}, H.~F. 2009, \icarus,
  199, 197

\bibitem[{{Lebreton} {et~al.}(2013){Lebreton}, {van Lieshout}, {Augereau},
  {Absil}, {Mennesson}, {Kama}, {Dominik}, {Bonsor}, {Vandeportal}, {Beust},
  {Defr{\`e}re}, {Ertel}, {Faramaz}, {Hinz}, {Kral}, {Lagrange}, {Liu}, \&
  {Th{\'e}bault}}]{2013A&A...555A.146L}
{Lebreton}, J., {van Lieshout}, R., {Augereau}, J.-C., {et~al.} 2013, \aap,
  555, A146

\bibitem[{{Levison} \& {Duncan}(1994)}]{1994Icar..108...18L}
{Levison}, H.~F. \& {Duncan}, M.~J. 1994, \icarus, 108, 18

\bibitem[{{Mamajek}(2012)}]{2012ApJ...754L..20M}
{Mamajek}, E.~E. 2012, \apjl, 754, L20

\bibitem[{{Mamajek} {et~al.}(2013){Mamajek}, {Bartlett}, {Seifahrt}, {Henry},
  {Dieterich}, {Lurie}, {Kenworthy}, {Jao}, {Riedel}, {Subasavage}, {Winters},
  {Finch}, {Ianna}, \& {Bean}}]{2013AJ....146..154M}
{Mamajek}, E.~E., {Bartlett}, J.~L., {Seifahrt}, A., {et~al.} 2013, \aj, 146,
  154

\bibitem[{{Marengo} {et~al.}(2009){Marengo}, {Stapelfeldt}, {Werner}, {Hora},
  {Fazio}, {Schuster}, {Carson}, \& {Megeath}}]{2009ApJ...700.1647M}
{Marengo}, M., {Stapelfeldt}, K., {Werner}, M.~W., {et~al.} 2009, \apj, 700,
  1647

\bibitem[{{Moons} \& {Morbidelli}(1993)}]{1993CeMDA..57...99M}
{Moons}, M. \& {Morbidelli}, A. 1993, Celestial Mechanics and Dynamical
  Astronomy, 57, 99

\bibitem[{{Morbidelli} \& {Moons}(1995)}]{1995Icar..115...60M}
{Morbidelli}, A. \& {Moons}, M. 1995, \icarus, 115, 60

\bibitem[{{Mustill} \& {Wyatt}(2012)}]{2012MNRAS.419.3074M}
{Mustill}, A.~J. \& {Wyatt}, M.~C. 2012, \mnras, 419, 3074

\bibitem[{{Quillen}(2006)}]{2006MNRAS.372L..14Q}
{Quillen}, A.~C. 2006, \mnras, 372, L14

\bibitem[{{Rodigas} {et~al.}(2013){Rodigas}, {Malhotra}, \&
  {Hinz}}]{2013arXiv1311.1207R}
{Rodigas}, T.~J., {Malhotra}, R., \& {Hinz}, P.~M. 2013, ArXiv e-prints

\bibitem[{{Shannon} {et~al.}(2014){Shannon}, {Clarke}, \&
  {Wyatt}}]{2014MNRAS.442..142S}
{Shannon}, A., {Clarke}, C., \& {Wyatt}, M. 2014, \mnras, 442, 142

\bibitem[{{Tamayo}(2014)}]{2014MNRAS.438.3577T}
{Tamayo}, D. 2014, \mnras, 438, 3577

\bibitem[{{van Leeuwen}(2007)}]{2007ASSL..350.....V}
{van Leeuwen}, F., ed. 2007, Astrophysics and Space Science Library, Vol. 350,
  {Hipparcos, the New Reduction of the Raw Data}

\bibitem[{{Wisdom}(1980)}]{1980AJ.....85.1122W}
{Wisdom}, J. 1980, \aj, 85, 1122

\bibitem[{{Wisdom}(1983)}]{1983Icar...56...51W}
{Wisdom}, J. 1983, \icarus, 56, 51

\bibitem[{{Wyatt} {et~al.}(1999){Wyatt}, {Dermott}, {Telesco}, {Fisher},
  {Grogan}, {Holmes}, \& {Pi{\~n}a}}]{1999ApJ...527..918W}
{Wyatt}, M.~C., {Dermott}, S.~F., {Telesco}, C.~M., {et~al.} 1999, \apj, 527,
  918

\bibitem[{{Yoshikawa}(1989)}]{1989A&A...213..436Y}
{Yoshikawa}, M. 1989, \aap, 213, 436

\end{thebibliography}

\end{document}